\documentclass{ut-thesis}
\usepackage[usenames]{color}
\usepackage{graphicx}
\usepackage{epsfig}
\usepackage{amssymb}
\usepackage{amsmath}
\usepackage{amsfonts}

\degree{Master of Science}
\department{Computer Science}
\gradyear{2007}
\author{Oktie Hassanzadeh}
\title{Benchmarking Declarative Approximate Selection Predicates}

\begin{document}

\begin{preliminary}

\maketitle



\begin{abstract}

\ignore{
Declarative data quality has been an active research topic.
The fundamental principle behind a declarative approach to
data quality is the use of declarative statements to realize data quality primitives
on top of any relational data source. A primary advantage of such an approach
is the ease of use and integration with existing applications.

Over the last couple of years several similarity predicates have been
proposed for common quality primitives (approximate selections, joins, etc)
and have been fully expressed using declarative SQL statements. In this
thesis new similarity predicates are proposed along with their declarative
realization, based on notions of probabilistic information retrieval. 
In particular it is shown how language models and hidden Markov models can
be utilized as similarity predicates for data quality and their
full declarative instantiation is presented. Also, it is shown how other scoring methods
from information retrieval, can be utilized in a similar setting.
Then, full declarative specifications of previously proposed
similarity predicates in the literature are presented, grouped into classes
according to their primary characteristics. 
Finally, a thorough performance and accuracy study comparing
a large number of similarity predicates for data cleaning operations is performed.
Their runtime performance as well as their accuracy are quantified
for several types of common quality problems encountered in operational
databases. 
}

Declarative data quality has been an active research topic.
The fundamental principle behind a declarative approach to
data quality is the use of declarative statements to realize data quality primitives
on top of any relational data source. A primary advantage of such an approach
is the ease of use and integration with existing applications.

Over the last couple of years several similarity predicates have been
proposed for common quality primitives (approximate selections, joins, etc.)
and have been fully expressed using declarative SQL statements. In this
thesis, new similarity predicates are proposed along with their declarative
realization, based on notions of probabilistic information retrieval. 
Then, full declarative specifications of previously proposed
similarity predicates in the literature are presented, grouped into classes
according to their primary characteristics. 
Finally, a thorough performance and accuracy study comparing
a large number of similarity predicates for data cleaning operations is performed.

\end{abstract}







\ignore{
\newpage

\begin{acknowledgements}
First, I would like to thank my supervisor, Nick Koudas. Special thanks to John Mylopoulos, the second reader of my thesis, for his valuable time and comments.

During my research, I had the pleasure of working in a wonderful atmosphere in the database lab. I had an unforgettable year with my colleagues there. While enjoying the taste of fresh coffee from our fancy coffee machine that helped us stay awake all long nights before the deadlines, we had many fruitful discussions that often resulted in brilliant new ideas. I would like to thank all my friends in the database lab, particularly Dimitris Tsirogiannis, Mohammad Sadoghi, Nilesh Bansal, Amit Chandel, Chaitanya Mishra, Nikos Sarkas and Manos Papagelis.

Finally, I would like to thank my parents and my brother for their love, support and guidance. While their physical absence in the last year and a half has truly been painful to me, their presence in my heart and in my mind has been and will always be the most valuable aspect of my life.

\end{acknowledgements}
}

\tableofcontents

\ignore{
\contentsline {chapter}{\numberline {1}Introduction}{1}
\contentsline {chapter}{\numberline {2}Related Work}{4}
\contentsline {chapter}{\numberline {3}Framework}{6}
\contentsline {section}{\numberline {3.1}Overlap Predicates}{7}
\contentsline {section}{\numberline {3.2}Aggregate Weighted Predicates}{8}
\contentsline {subsection}{\numberline {3.2.1}Tf-idf Cosine Similarity}{8}
\contentsline {subsection}{\numberline {3.2.2}BM25 Predicate}{9}
\contentsline {section}{\numberline {3.3}Language Modeling Predicates}{9}
\contentsline {subsection}{\numberline {3.3.1}Language Modeling}{10}
\contentsline {subsection}{\numberline {3.3.2}Hidden Markov Models}{11}
\contentsline {section}{\numberline {3.4}Edit-based Predicates}{11}
\contentsline {section}{\numberline {3.5}Combination Predicates}{12}
\contentsline {chapter}{\numberline {4}Declarative Framework}{14}
\contentsline {section}{\numberline {4.1}Overlap Predicates}{15}
\contentsline {section}{\numberline {4.2}Aggregate Weighted Predicates}{16}
\contentsline {subsection}{\numberline {4.2.1}Tf-idf Cosine Similarity}{16}
\contentsline {subsection}{\numberline {4.2.2}BM25}{17}
\contentsline {section}{\numberline {4.3}Language Modeling Predicates}{17}
\contentsline {subsection}{\numberline {4.3.1}Language Modeling}{17}
\contentsline {subsection}{\numberline {4.3.2}Hidden Markov Models}{20}
\contentsline {section}{\numberline {4.4}Edit-based Predicates}{20}
\contentsline {section}{\numberline {4.5}Combination Predicates}{21}
\contentsline {chapter}{\numberline {5}Evaluation}{26}
\contentsline {section}{\numberline {5.1}Benchmark}{26}
\contentsline {section}{\numberline {5.2}Evaluating Accuracy}{30}
\contentsline {section}{\numberline {5.3}Settings}{31}
\contentsline {subsection}{\numberline {5.3.1}Choice of Weights for Weighted Overlap Predicates}{31}
\contentsline {subsection}{\numberline {5.3.2}Parameter Settings for Predicates}{32}
\contentsline {subsection}{\numberline {5.3.3}Q-gram Generation}{32}
\contentsline {section}{\numberline {5.4}Accuracy Results}{33}
\contentsline {subsection}{\numberline {5.4.1}Comparison of predicates}{35}
\contentsline {section}{\numberline {5.5}Performance Results}{37}
\contentsline {subsection}{\numberline {5.5.1}Preprocessing}{37}
\contentsline {subsection}{\numberline {5.5.2}Query time}{38}
\contentsline {subsection}{\numberline {5.5.3}Scalability}{39}
\contentsline {section}{\numberline {5.6}Performance Enhancements}{40}
\contentsline {section}{\numberline {5.7}Summary of Evaluation}{42}
\contentsline {chapter}{\numberline {6}Conclusions}{48}
\contentsline {chapter}{Bibliography}{48}
\contentsline {chapter}{\numberline {A}Data preparation SQL Statements}{53}
\contentsline {section}{\numberline {A.1}Qgram generation}{53}
\contentsline {section}{\numberline {A.2}Word token generation}{54}
\contentsline {section}{\numberline {A.3}Qgram generation of the word tokens (for combination predicates)}{54}
\contentsline {chapter}{\numberline {B}SQL Statements for Predicates}{55}
\contentsline {section}{\numberline {B.1}Overlap Predicates}{55}
\contentsline {subsection}{\numberline {B.1.1}IntersectSize}{55}
\contentsline {subsection}{\numberline {B.1.2}Jaccard}{56}
\contentsline {subsection}{\numberline {B.1.3}WeightedMatch}{57}
\contentsline {subsection}{\numberline {B.1.4}WeightedJaccard}{58}
\contentsline {section}{\numberline {B.2}Aggregate Weighted Predicates}{59}
\contentsline {subsection}{\numberline {B.2.1}Tfidf Cosine Predicate}{59}
\contentsline {subsection}{\numberline {B.2.2}BM25 Predicate}{61}
\contentsline {section}{\numberline {B.3}Language Modeling Predicates}{62}
\contentsline {subsection}{\numberline {B.3.1}Language Modeling}{62}
\contentsline {subsection}{\numberline {B.3.2}Hidden Markov Models}{64}
\contentsline {section}{\numberline {B.4}Combination Predicates}{65}
\contentsline {subsection}{\numberline {B.4.1}GES$^{Jaccard}$}{65}
\contentsline {subsection}{\numberline {B.4.2}GES$^{apx}$}{67}
\contentsline {subsection}{\numberline {B.4.3}SoftTFIDF}{69}
}




\end{preliminary}



\chapter{Introduction}

The importance of data cleaning and quality technologies for business practices
is well recognized. Data cleaning has been an active research topic in several
communities including statistics, machine learning and data management. 
The quality of data suffers from typing mistakes, lack of standards for
recording database fields, integrity constraints that are not enforced, inconsistent
data mappings, etc. For years, data quality technology has grown independently from core data 
management. Data quality tools became part of Extract Transform Load (ETL)
technologies, commonly applied during the initial loading phase of
data into a warehouse.
Although this might be a viable approach for data analytics,
where data processed are static, it is far from acceptable for operational databases.
Dynamic databases however, face proliferating quality problems, that degrade
common business practices. 

Recently, there has been a major focus on tighter integration of data quality technology
with database technology. In particular there has been research work on the
efficient realization of popular data cleaning algorithms inside database
engines as well as studies for the efficient realization of data quality primitives
in a declarative way. The approaches are complementary, the former assuring
great performance and the latter ease of deployment and integration with
existing applications without modification of the underlying database engine.
We are concerned with declarative 
implementations of data quality primitives in this thesis.
In particular we study declarative realizations of several {\em similarity predicates}
for the popular approximate (flexible) selection 
operation for data de-duplication \cite{Koudas05, Koudas06}. 
A similarity predicate $sim()$ is a predicate that numerically quantifies
the 'similarity' or 'closeness' of two (string) tuples.
Given a relation $R$, the approximate selection operation 
using similarity predicate $sim()$, 
will report all tuples $t \in R$ such that
$sim(t_q,t) \geq \theta$, where $\theta$ a specified numerical 
'similarity threshold' and $t_q$ a query tuple. Approximate selections
are special cases of the approximate join (record linkage, similarity join) 
operation \cite{Koudas05, Koudas06}.
Several efficient declarative implementations of this operation for specific
similarity predicates have been proposed \cite{Koudas05, Koudas06} both for approximate selections
and joins.

In this thesis, we conduct a thorough study of declarative realizations of
similarity predicates for approximate selections.  We introduce and adapt
novel predicates, realize them declaratively and compare them with existing
ones for accuracy and performance. In particular we make the following
contributions:
\begin{itemize}
\item
Inspired by the success of {\em tf-idf cosine similarity} from information retrieval \cite{Salton88}
as a similarity predicate for approximate selections, we introduce declarative
realizations of other successful predicates from information retrieval and in
particular the popular BM25 measure.
\item
We introduce declarative realizations of probabilistic similarity predicates inspired by
Language Models from information retrieval \cite{ponte98} and Hidden Markov Models \cite{hmmSIGIR99},
suitably adapted for the case of approximate selections.
\item
We present declarative realizations of previously proposed similarity predicates
for the approximate selection problem and we propose a categorization of all measures
both previously proposed and new according to their characteristics.
\item
We present a thorough experimental study comparing all similarity predicates
for accuracy and performance, under various types of quality problems in the
underlying data.
\end{itemize}

\chapter{Related Work}

Data quality has been an active research topic for many years.
A collection of statistical techniques have been introduced
initially for the record linkage problem \cite{Fellegi69, linkage90}. 
The bulk of early work on data quality was geared towards correcting problems in
census files \cite{Winkler99}. A number of similarity predicates were developed
taking into account the specific application domain (i.e., census files) for assessing
closeness between person names (e.g., Jaro, Jaro-Winkler \cite{Jaro84,Winkler99}, etc).

The work of Cohen \cite{Cohen98} introduced the use of primitives from
information retrieval (namely cosine similarity, utilizing tf-idf\cite{Salton88}) to identify
flexible matches among database tuples. 
A performance/accuracy study conducted by Cohen et al., \cite{cohen1} demonstrated
that such techniques outperform common predicates introduced for specific
domains (e.g., Jaro, Jaro-Winkler, etc). 

Other techniques geared towards
database tuples include the merge/purge technique \cite{MergePurge}. Several
predicates to quantify approximate match between strings have been utilized
for dealing with quality problems, including edit distance and its variants
\cite{GusfieldBook}. Hybrid predicates combining notions of edit distance and
cosine similarity have also been introduced \cite{fmsSIGMOD03, ananthakrishna02eliminating}.
Recently, \cite{SSJOIN,SSJOIN2} presented SSJOIN, 
a primitive operator for efficient
set similarity joins. Utilizing ideas from \cite{sunitaSIGMOD04}, such
an operator can be used for approximate matching based on a
number of similarity functions, including hamming distance, edit-distance and
Jaccard similarity. However, the choice of the similarity predicate in this approach is limited \cite{SSJOIN2}.
The bulk of the techniques and predicates however have been introduced without a
declarative framework in mind. Thus, integrating them with applications
utilizing databases in order to enable approximate selections  is not very easy.

Gravano et al. \cite{joinForFreeVLDB01, Galhardas01},
 introduced a declarative methodology
for realizing approximate joins and selections for edit distance. Subsequently
a declarative framework for realizing tf-idf cosine similarity was introduced
\cite{textjoinWWW03,flexiblematchVLDB04, Koudas06Spider, SpiderSIGMOD05}.

There has been a great deal of research in the information
retrieval literature on weighting schemes beyond cosine similarity with
tf-idf weighting.
Recent IR research has shown BM25 to be the most effective among the known
weighting schemes \cite{bm25TREC95}. This weighting scheme models
the distribution of within-document term frequency, document length and query
term frequency very accurately. 
Moreover, in the information retrieval literature, language modeling has
been a very active research topic as an alternate scheme to weight documents
for their relevance to user queries. Starting with Ponte and Croft \cite{ponte98}
language models for information retrieval have been widely studied.

Hidden Markov Models (HMM) have been very successful in machine learning
and they have been utilized for a variety of learning tasks such as
named entity recognition and voice recognition\cite{rabiner89}.
They have also been utilized for information retrieval as
well \cite{hmmSIGIR99}. An experimental study on
TREC data demonstrated that an extremely simple realization of HMM
outperforms standard tf-idf for information retrieval \cite{hmmSIGIR99}.
Several researchers \cite{robertson04} have tried to formally reason
about the relative goodness of information retrieval weighting schemes.

\chapter{Framework}
\label{framework}

Let $Q$ be a query string and $D$ a string record from a base
relation $R=\{D_i: 1 \le i \le N\}$. We denote by $\cal{Q}$, $\cal{D}$ 
the set of {\em tokens} in $Q$ and $D$ respectively. We refer to substrings of
a string as tokens in a generic sense. Such tokens can be words or
q-grams (sequence of $q$ consecutive characters of a string) for example. 
For $Q$=`db lab', $\cal{Q}$=$\{$`db', `lab'$\}$ for word-based
tokenization and $\cal{Q}$=$\{$`db ' ,`b l',` la', `lab'$\}$ for tokenization
using 3-grams. We refer to tokens throughout the thesis when referring to words
or q-grams. We make the choice specific (word or q-gram) for 
techniques we present, when is absolutely required. In certain
cases, we may associate a {\em weight} with each token. Several weighting
mechanisms exist. We present our techniques referring to weights of
tokens, making the choice of the weighting scheme concrete when required. 
In chapter \ref{experimental}
we realize our techniques for specific choice of tokens and specific weighting
mechanisms.

Our goal is to calculate a {\em similarity score} 
between $Q$ and $D$ using a similarity predicate. We group similarity
predicates into five classes based on their characteristics, namely:
\begin{itemize}
\item {\bf Overlap predicates:} These are predicates that assess similarity
based on the overlap of tokens in $\cal{Q},\cal{D}$.
\item {\bf Aggregate Weighted Predicates:} Predicates that assess
similarity by manipulating weights (scores) assigned to elements of
$\cal{Q},\cal{D}$
\item {\bf Language Modeling Predicates:} Predicates that are based on
probabilistic models imposed on elements of $\cal{Q},\cal{D}$
\item {\bf Edit Based Predicates:} Predicates based on a set of edit operations 
applied between $Q$ and $D$.
\item {\bf Combination Predicates:} Predicates combining features from the
classes above.
\end{itemize}
The classes were defined by studying the properties of previously proposed
similarity predicates as well as ones newly proposed herein. 
The first four classes encompass predicates introduced previously
in various contexts for data cleaning tasks, with the exception of
BM25 which to the best of our knowledge is the first time
that is deployed for data cleaning purposes. The Language Modeling class of
predicates draws from work on information retrieval and is introduced herein for
data cleaning tasks. Within each class we discuss declarative realizations
of predicates.

\section{Overlap Predicates}

Suppose $\cal{Q}$  is the set of tokens in the query string $Q$ and $\cal{D}$
is the set of tokens in the string tuple $D$. The {\em IntersectSize} predicate
\cite{sunitaSIGMOD04} is simply the
number of common tokens between $Q$ and $D$, i.e.:
\begin{equation}
 sim_{intersect}(Q,D) =  |\cal{Q} \cap \cal{D}|
\end{equation}
Jaccard similarity \cite{sunitaSIGMOD04} 
is the fraction of tokens in $Q$ and $S$ that are present
in both, namely:
\begin{equation}
 sim_{Jaccard}(Q,D) = \frac {|\cal{Q} \cap \cal{D}|}{ |\cal{Q} \cup \cal{D}| }
\end{equation}
If we assign a weight $w(t)$\footnote{Discussion of ways to
assign such weights to tokens follows in subsequent chapters.} 
to each token $t$, we can define weighted
versions of the above predicates. {\em WeightedMatch} \cite{sunitaSIGMOD04} is the total weight of common
tokens in $\cal{Q}$ and $\cal{D}$, i.e., $\sum_{t \in \cal{Q} \cap
\cal{D}}{w(t)}$. Similarly, {\em WeightedJaccard} is the sum of the weights of
tokens in $|\cal{Q} \cap \cal{D}|$ divided by the sum of
the weights of tokens in $|\cal{Q} \cup \cal{D}|$.

\section{Aggregate Weighted Predicates}

The predicates in this class encompass predicates widely
adopted from information retrieval (IR). 
A basic task in IR is, given a query, identifying \textit{relevant}
\textit{documents} to that query. In  our context, we would like to identify
the \textit{tuples} in a relation that are \textit{similar} to a query string.

Given a query string $Q$ and a string tuple $D$, the similarity score of $Q$
and $D$ in this class of predicates  is of the form $sim(Q,D) = \sum_{t \in \cal{Q
\cap D}} w_{q}(t,Q) w_{d}(t,D)$, where $w_q(t,Q)$ is the query-based weight of
the token $t$ in string $Q$ and $w_d(t,D)$ is the tuple-based weight of
the token $t$ in string $D$.

\subsection{Tf-idf Cosine Similarity}
The tf-idf cosine similarity\cite{Salton88} between
a query string $Q$ and a string tuple $D$ is defined as follows:
\ignore{\begin{eqnarray}
  sim_{cosine}(Q,D) &=& \sum_{t \in \cal{Q \cap  D}} w_{q}(t,Q) w_{d}(t,D)
\end{eqnarray}
}
\begin{eqnarray}
  sim_{cosine}(Q,D) &=& \sum_{t \in \cal{Q \cap  D}} w_q(t,Q) w_d(t,D)
\end{eqnarray}
where $w_q(t,Q),w_d(t,D)$ are the normalized tf-idf weights \cite{Salton88}.
The normalized tf-idf between a token $t$ and a string $S$, $w(t,S)$ is given by:
\begin{eqnarray}
\nonumber  w(t,S) = \frac{w'(t,S)}{\sqrt{ \sum_{{t' \in  \cal{S}}}w'(t',S)^2} } &,&
w'(t,S) = tf(t,S) . idf(t)
\ignore{\nonumber  \displaystyle w_q(t,Q) = \frac{w'_Q(t)}{\sum_{{t' \in 
Q}}w'_Q(t')^2} &,& \displaystyle w'_Q(t) = TF_Q(t) IDF(t) }
\end{eqnarray}
The $idf$ term makes the weight of a token inversely
proportional to its frequency in the database; the $tf$ term makes it
proportional to its frequency in $S$.  Intuitively, this
assigns low scores to frequent tokens and high scores to rare
tokens. More discussion is available elsewhere \cite{Cohen98,textjoinWWW03}.

\subsection{BM25 Predicate}

The $BM25$ similarity score between a query string $Q$ and a tuple $D$, is given
as:
\begin{equation}
  sim_{BM25}(Q,D) =  \sum_{t \in \cal {Q \cap  D}} w_{q}(t,Q) w_{d}(t,D)
\end{equation} 
where
\begin{eqnarray}
\nonumber w_{q}(t,Q) & = & \frac{(k_3+1)*tf(t,Q)}{k_3 + tf(t,Q)}  \\
\nonumber w_{d}(t, D) & = & w^{(1)}(t,R) \frac{(k_1+1)*tf(t,D)}{K(D) + tf(t,D)}  
\end{eqnarray} 
$w^{(1)}$ is a modified form of Robertson-Sparck Jones weight:
\begin{eqnarray} 
\label{rs_wt}
w^{(1)}(t,R) & = & log \left( \frac{N - n_t +0.5}{ n_t + 0.5} \right) \\
\nonumber K(D) & = & k_1 \left( (1-b) + b \frac{\vert D \vert}{ avgdl }
\right) 
\end{eqnarray} 
 and $N$ is the number  of tuples in the base relation $R$, $n_t$ is the number of
tuples in $R$ containing the token $t$, $tf(t,D)$ is the frequency of occurrence of
the token $t$ within tuple $D$, $\vert D \vert$ is the number of
tokens of tuple $D$, $avgdl$ is the average number of tokens
per tuple, i.e.
$\frac{\sum_{D \in R}\vert D \vert}{N}$ and $k_1$, $k_3$, and $b$ are
independent parameters. For TREC-4 experiments
\cite{bm25TREC95}, $k_1 \in [1,2]$, $k_3=8$ and $b\in [0.6, 0.75]$.

\section{Language Modeling Predicates}

A language model, is a form of a probabilistic model. To realize
things concretely, we base our discussion
on a specific model introduced by Ponte and Croft \cite{ponte98}. Given
a collection of documents, a language
model is inferred for each; then the probability of generating a given query
according to each of these models is estimated and documents are ranked
according to these probabilities. Considering an approximate
selection query,  each tuple in the database is considered 
as a document; a model is inferred for each tuple
and the probability of generating the query given the
model is the similarity between the query and the tuple.

\subsection{Language Modeling}
 
The similarity score between query $Q$ and tuple $D$ is defined as:
\begin{equation}
\label{eqlm1}
\displaystyle
 sim_{LM}(Q,D) = \hat{p}(Q|M_D) = \prod_{t \in \cal{Q}} \hat{p}(t|M_D) \times
\prod_{t \notin \cal{Q}}(1 - \hat{p}(t|M_D))
\end{equation}
 where $\hat p(t|M_D)$ is the probability of token $t$ occurring in tuple $D$ and is given as follows:
 \begin{equation}
 \label{eqlm2}
 \displaystyle \hat p(t|M_D) = \begin{cases}
                               \hat p_{ml}(t,D)^{(1.0-\hat{R}_{t,D})} \times \hat p_{avg}(t)^{\hat{R}_{t,D}}
                                  & \text{if $tf_{(t,D)} > 0$}\\
                               \frac{cf_t}{cs} & \text{otherwise}
                               \end{cases}
 \end{equation}
  $\hat p_{ml}(t,D)$ is  the maximum likelihood estimate of the probability of
the token $t$ under the token distribution for tuple $D$ and is equal to
$\frac{tf_{(t,D)}} {dl_D} $ where $tf_{(t,D)}$ is raw term frequency and $dl_D$
is the total number of tokens in tuple $D$. 
$\hat p_{avg}(t)$ is the mean probability of token $t$ in documents containing it, i.e.,
 \begin{equation}
 \label{eqlm3}
 \displaystyle \hat p_{avg}(t) = \frac{(\sum_{D_{(t \in D)}} {\hat p_{ml}(t|M_D)} )} {df_t}
 \end{equation}
  where $df_{t}$ is the  document frequency of token $t$. This term is used
since we only have a tuple sized sample from the distribution of $M_D$, thus the
maximum likelihood estimate is not reliable enough; we need an estimate from 
a larger amount of data. The term $\hat R_{t,d}$  is used to model the risk for a 
term $t$ in a document $D$ using a geometric distribution:
 \begin{equation}
 \label{eqlm4}
 \displaystyle \hat R_{t,D} = \left(\frac{1.0}{(1.0+\bar f_{t,D})}\right) \times 
                              \left(\frac{\bar f_{t,D}}{(1.0+\bar f_{t,D})}\right)^{tf_{t,D}}
 \end{equation}
 $\bar f_{t,D}$ is the expected term count for token $t$ in 
tuple $D$ if the token occurred at the average rate, i.e., $p_{avg}(t) \times dl_D$. 
 The intuition behind this formula is that as the $tf$ gets further away from the
normalized mean, the mean probability becomes riskier to use as an estimate. %
\ignore{ \textbf{Note.} \cite{ponte98} uses notation $\bar f_t$ although its value
also depends on the size of the document $d$, so we use $\bar{f}_{t,d}$ instead.}
Finally, $cf_t$ is the raw count of token $t$ in the collection, i.e.
$\sum_{D \in R}tf(t,D)$ and $cs$ is the
raw collection size or the total number of tokens in the collection, i.e.
$\sum_{D \in R} dl_D$.
$\frac{cf_t}{cs}$ is used as the probability of observing a non-occurring token.

\subsection{Hidden Markov Models}

\begin{figure}
\centering
\epsfig{file=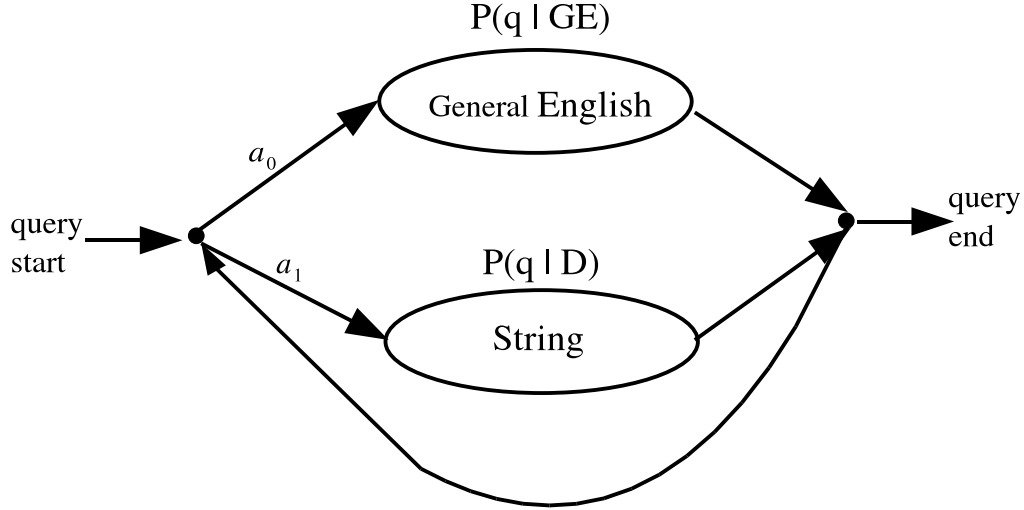,width=5.3in}
\caption{Two State Hidden Markov Model}
\label{hmm-fig}
\end{figure}

The query generation process can be modeled by a discrete Hidden Markov process.
Figure \ref{hmm-fig} shows a simple yet powerful two-state HMM for this process. The first
state, labeled ``String" represents the choice of a token directly from the string.
The second state, labeled ``General English" represents the choice of a token that
is unrelated to the string, but occurs commonly in queries.

Suppose $Q$ is the query string and $D$ is a string tuple from the base
relation $R$; the similarity score between $Q$ and $D$, $sim_{HMM}(Q,D)$, is equal to the probability of
generating $Q$ given that $D$ is similar, that is:
 \begin{equation}  \label{eqhmm1} 
 \displaystyle P(Q|D\text{ is similar}) = \prod_{q \in \cal{Q}} ({a_0 P(q|GE) +
a_1
P(q|D))}
 \end{equation}
 where:
 \begin{equation}
 \displaystyle P(q|D) = \frac{\text{number of times $q$ appears in $D$}}
{\text{length of $D$}}
 \end{equation}
 \begin{equation}
 P(q|GE) = \frac{\sum_{D \in R}{\text{number of times $q$ appears in $D$}}}
{\sum_{D \in R}{\text{length of $D$}}}
 \end{equation} 
and $a_0$ and $a_1 = 1 - a_0$ are transition probabilities of the HMM.  The
values for these parameters can be optimized to maximize accuracy given training
data.

\section{Edit-based Predicates}

An important and widely used class of string matching predicates is the
class of edit-based predicates. In this class, the similarity between $Q$ and
$D$ is the transformation cost of string $Q$ to $D$, 
$tc(Q,D)$. More specifically $tc(Q,D)$ is defined as the
minimum cost sequence of edit operations that converts $Q$ to $D$. Edit operations
include {\em copy, insert, substitute} and {\em delete} characters in $Q$ and
$D$ \cite{GusfieldBook}. Algorithms exist to compute $tc(Q,D)$ in polynomial
time \cite{GusfieldBook} but complexity is sensitive to
the nature of operations and their operands (individual characters, blocks of
consecutive characters, etc). The edit similarity is therefore defined as:
   \begin{equation}
 sim_{edit}(Q,D) = 1 - \frac{tc(Q,D)}{\max \{|Q|, |D|\}}
 \end{equation}
Edit operations have an associated cost. 
In the Levenstein edit-distance \cite{GusfieldBook} which we will refer to 
as edit-distance,  the cost of copy operation is zero and all other 
operations have unit cost. Other cost models are also possible \cite{GusfieldBook}.

\section{Combination Predicates}

We present a general similarity predicate and refer to it as generalized edit
similarity (GES) (following \cite{SSJOIN}). Consider two strings $Q$
and $D$ that are tokenized into {\em word} tokens and a weight function $w(t)$ that 
assigns a weight to each word token $t$. The transformation cost of
string $Q$ to $D$, $tc(Q,D)$ is the minimum cost of transforming $Q$ to $D$ by
a sequence of the following transformation operations:
\begin{itemize}
\item \textit{token replacement}: Replacing word token $t_1$ in $Q$ by word
token $t_2$ in $D$ with cost $[1-sim_{edit}(t_1,t_2)] \cdot w(t_1)$, where
$sim_{edit}(t_1,t_2)$ is the edit similarity score between $t_1$ and $t_2$. 
\item \textit{token insertion}: Inserting a word token $t$ into $Q$ with cost
$c_{ins} \cdot w(t)$ where $c_{ins}$, is a constant token insertion factor, with
values between 0 and 1.
\item \textit{token deletion}: Deleting a word token $t$ from $Q$ with cost
$w(t)$.
\end{itemize}
Suppose $wt(Q)$ is the sum of weights of all word tokens in the string $Q$. We
define the \textit{generalized edit similarity} predicate between a query
string $Q$ and a tuple $D$ as follows:
 \begin{equation} 
 \label{GES}
 \displaystyle sim_{GES}(Q,D) = 1 - \min\left( \frac{tc(Q,D)}{wt(Q)} , 1.0 \right)
 \end{equation}

A related predicate is the SoftTFIDF predicate \cite{cohen1}. 
In SoftTFIDF, normalized tf-idf weights of word tokens are used
along with cosine similarity and any other similarity function $sim(t,r)$ to find
the similarity between word tokens. Therefore the similarity score, 
$sim_{SoftTFIDF} (Q,D)$, is equal to:
\begin{equation}
 \label{softtfidf} \small
  \sum_{t \in CLOSE(\theta,Q,D)} w(t,Q) \cdot w( \arg\max_{r \in \cal{D}} (sim(t,r)), D ) \cdot \max_{r \in \cal{D}} (sim(t,r))
\end{equation}
 where $w(t,Q),w(t,D)$ are the normalized tf-idf weights and $CLOSE(\theta,Q,D)$ is the set of words $t \in \cal{Q}$ such that there
exists some $v \in \cal{D}$ such that $sim(t,v) > \theta$.

\chapter{Declarative Framework}
\label{declarative_chapter}

We now describe declarative realizations of predicates in each
class. We present declarative statements using standard SQL expressions. 
For all predicates, there is  a preprocessing phase responsible for
 tokenizing strings in
the base relation, $R$, and calculating as well as storing related weight values which
are subsequently utilized at query time.
Tokenization of relation $R$ (\texttt{BASE\_TABLE}) creates the table
\texttt{BASE\_TOKENS} \texttt{(tid, token)}, where $tid$ is a unique
token identifier for each tuple of \texttt{BASE\_TABLE} and {\em token} an
associated token (from the set of tokens corresponding to the tuple with
identifier $tid$ in \texttt{BASE\_TABLE}).
The query string is also tokenized on the fly (at query time) creating 
the table \texttt{QUERY\_TOKENS(token)}.
 
In the rest of this chapter, we present SQL expressions required for
preprocessing and query time approximate selections for the different predicates.
In some cases, we re-write formulas to make them amenable to
more efficient declarative realization. The main SQL codes are given along with their description here. Appendix A
contains detailed SQL expressions.

\section{Overlap Predicates}

The IntersectSize predicate requires token generation to be completed
in a preprocessing step. SQL statements to conduct such a tokenization,
which is common to all predicates we discuss, is available in Appendix A. The SQL
statement for approximate selections with the IntersectSize predicate
is shown on Figure \ref{intersect}. The Jaccard
coefficient predicate can be efficiently computed by storing the number of tokens for
each tuple of the \texttt{BASE\_TABLE} during the preprocessing step. For this
reason we create a table \texttt{BASE\_DDL(tid, token, len)} where \texttt{len} is the
number of tokens in tuple with tuple-id \texttt{tid}. The SQL statement
for conducting approximate selections with the Jaccard predicate
is presented in Figure \ref{jaccard}.
  
\begin{figure}
\centering
 \small \tt  
\begin{tabular}{|ll|} \hline
 \multicolumn{2}{|l|}{
INSERT INTO INTERSECT\_SCORES (tid, score)} \\
SELECT & R1.tid, COUNT(*) \\
FROM   & BASE\_TOKENS R1, QUERY\_TOKENS R2 \\
WHERE  & R1.token = R2.token \\
GROUP BY & R1.tid  \\ \hline
\end{tabular}
\normalfont
\caption{SQL Code for IntersectSize}
\label{intersect}
\end{figure}

\begin{figure}
\centering
 \small \tt 
\begin{tabular}{|ll|} \hline
\ignore{
\multicolumn{2}{|c|}{Preprocessing} \\  \hline
\multicolumn{2}{|l|}{INSERT INTO BASE\_LENGTH (tid, len)} \\
SELECT   & T.tid, COUNT(*) \\
FROM     & BASE\_TOKENS T \\
GROUP BY & T.tid\\ \hline
\multicolumn{2}{|c|}{Query} \\  \hline
}
\multicolumn{2}{|l|}{INSERT INTO JACCARD\_SCORES (tid, score)}\\
SELECT   & S1.tid, COUNT(*)/(S1.len+S2.len-COUNT(*))  \\
FROM     & BASE\_DDL S1, QUERY\_TOKENS R2, \\
         & (SELECT COUNT(*) AS len \\ 
         & \hspace{0.02in} FROM QUERY\_TOKENS) S2 \\
WHERE    & S1.token = R2.token \\
GROUP BY & S1.tid, S1.len, S2.len \\ \hline

\end{tabular}
\normalfont
\caption{SQL Code for Jaccard Coefficient}
\label{jaccard}
\end{figure}

The weighted overlap predicates require
calculation and storage of the related weights for tokens of the base
relation during preprocessing. For the WeightedMatch predicate, 
we store during the preprocessing step the weight of
each token redundantly with each \texttt{tid, token} pair in a table
\texttt{BASE\_TOKENS\_WEIGHTS(tid, token, weight)} in order to avoid an extra join
with a table \texttt{BASE\_WEIGHT(token, weight)} at query time. 
In order to calculate the similarity score at query time, 
we use SQL statements similar to that used for the 
IntersectSize predicate (shown in Figure \ref{intersect}) but replace
table \texttt{BASE\_TOKENS} by \texttt{BASE\_TOKENS\_WEIGHTS} and \texttt{COUNT(*)}, by \texttt{SUM(R1.weight)}.

For the WeightedJaccard predicate, we create during preprocessing a table 
\texttt{BASE\_DDL(tid, token, weight, len)}   where \texttt{weight} is the
weight of \texttt{token} and \texttt{len} is the sum of weights of
tokens in the tuple with tuple-id \texttt{tid}. The SQL statement for
approximate selections using this predicate is the same as the one 
shown in Figure \ref{jaccard} but \texttt{COUNT(*)} is replaced by \texttt{SUM(weight)}.

\ignore{
The SQL code that stores $IDF$ weights in
relation \texttt{BASE\_IDF(token, idf)} during the preprocessing time
is presented in Figure \ref{idf}.

\begin{figure}
\centering
 \small \tt  
\begin{tabular}[ht]{|ll|} \hline
\multicolumn{2}{|l|}{INSERT INTO BASE\_SIZE} \\
SELECT & COUNT(*) \\
FROM   & BASE\_TABLE \\
& \\
\multicolumn{2}{|l|}{INSERT INTO BASE\_IDF} \\
SELECT   & T.token, LOG(S.size)-LOG(COUNT(DISTINCT T.tid)) \\
FROM     & BASE\_TOKENS T, BASE\_SIZE S \\
GROUP BY & T.token \\
\hline
\end{tabular}
\normalfont
\caption{SQL Code for idf Weights Calculation}
\label{idf}
\end{figure}
}

\section{Aggregate Weighted Predicates}

\subsection{Tf-idf Cosine Similarity}

The SQL implementation of the tf-idf cosine similarity predicate has been presented in
\cite{textjoinWWW03}. During preprocessing, we store 
tf-idf weights for the base relation in relation 
\texttt{BASE\_WEIGHTS(tid, token, weight)}.
A weight table \texttt{QUERY\_WEIGHTS(token, weight)} for the 
query string is created on the
fly at query time. The SQL statements in
Figure \ref{awsql} will calculate the similarity score for each tuple of the 
base table.

\begin{figure}
\centering
 \small \tt
\begin{tabular}{|ll|} \hline
\multicolumn{2}{|l|}{INSERT INTO SIM\_SCORES (tid, score)} \\
SELECT   & R1W.tid, SUM(R1W.weight*R2W.weight)\\
FROM     & BASE\_WEIGHTS R1W, QUERY\_WEIGHTS R2W \\
WHERE    & R1W.token = R2W.token \\
GROUP BY & R1W.tid \\ \hline
\end{tabular}
\normalfont
\caption{SQL Code for Aggregate Weighted Predicates}
\label{awsql}
\end{figure}

\subsection{BM25}

Realization of BM25 in SQL involves generation of the table \texttt{BASE\_WEIGHTS(tid, token, weight)} 
storing the weights for tokens in each tuple of the base relation. These weights ($w_d(t,D)$) consist of two parts that could be considered as modified versions of tf and idf. For a complete set of
SQL statements implementing the required preprocessing, refer to Appendix A. 
The query weights table \texttt{QUERY\_WEIGHTS(token, weight)} can be created on the fly using the 
following subquery: \\
{ \tt \small 
(SELECT        TF.token, TF.tf*($k_3$+1)/($k_3$+TF.tf) AS weight \\
  \indent FROM    \indent ( SELECT T.token, COUNT(*) AS tf \\
  \indent \indent \indent \indent \indent \indent FROM      QUERY\_TOKENS T \\
  \indent \indent \indent \indent \indent \indent GROUP BY \hspace{0.05in} T.token ) TF) \\
}

 The SQL statement shown in Figure \ref{awsql} will calculate BM25 similarity scores.

\section{Language Modeling Predicates}
\label{lmdeclarative}

\subsection{Language Modeling}

In order to calculate language modeling scores efficiently, we rewrite the
formulas and finally drop some terms that would not affect the overall accuracy
of the metric.
Calculating the values in equations (\ref{eqlm3}) and (\ref{eqlm4}) is
easy. We build the following relations during preprocessing:
 \texttt{BASE\_TF(tid,token,tf)} where \texttt{tf}$=tf_{token,tid}$. \\
 \texttt{BASE\_DL(tid,dl)}  where \texttt{dl}$=dl_{tid}$.\\
  \texttt{BASE\_PML(tid,token,pml)} where
\texttt{pml}$=\hat p_{ml}= \frac{tf_{token,tid}}{dl_{tid}}$.\\
 \texttt{BASE\_PAVG(token,pavg)} where \texttt{pavg}$=\hat
p_{avg}(token)$.\\
 \texttt{BASE\_FREQ(tid,token,freq)} where \texttt{freq}$=\bar
f_{token,tid}$.\\
 \texttt{BASE\_RISK(tid,token,risk)} where \texttt{risk}$=\hat R_{token,tid}$.

We omit most of the SQL statements in this chapter for readability. Full SQL statements are available in Appendix A.
In order to improve the performance of the associated SQL queries, 
we rewrite the final score formula of equation (\ref{eqlm1}), as follows:

\begin{equation}
 \label{eqlm5}
 \begin {aligned} 
  \displaystyle
  \hat{p}(Q|M_D) & = \displaystyle \left[ \prod_{t \in \cal{Q}} \hat{p}(t|M_D)
\right]
   \times \frac{\displaystyle \prod_{\forall t}(1 - \hat{p}(t|M_D))}
{\displaystyle \prod_{t \in \cal{Q}}(1 - \hat{p}(t|M_D))}
 \end{aligned}
\end{equation}

 We slightly change (\ref{eqlm5}) to the following:
\begin{equation}
  \label{eqlm6}
\begin {aligned} 
\displaystyle
  \hat{p}(Q|M_D) & = \displaystyle \left[ \prod_{t \in \cal{Q}} \hat{p}(t|M_D)
\right]
 \times \frac{\displaystyle \prod_{\forall t \in \cal{D}}(1 - \hat{p}(t|M_D))}
{\displaystyle \prod_{t \in \cal{Q} \cap \cal{D}}(1 - \hat{p}(t|M_D))}
 \end{aligned}
\end{equation}

This change results in a large performance gain, since the computation is 
restricted to the tokens of the query and the tokens of a tuple (as opposed to
the entire set of tokens present in the base relation). Experiments demonstrate
that accuracy is not considerably affected.

In equation (\ref{eqlm2}), we only materialize the first part (i.e., values of
tokens that are present in the tuple $D$) in the relation \texttt{BASE\_PM}
during preprocessing (storing the second part
would result in unnecessary waste of space). We therefore have to divide all
formulas that use $\hat p(t|M_D)$ into two parts: one for tokens present in the
tuple under consideration and one for all other tokens. 
So we rewrite the first term in equation (\ref{eqlm6}) as follows:
\begin{equation}
\label{eqlm7}
\small
\begin {aligned} 
\prod_{t \in \cal{Q}} \hat{p}(t|M_D) & = \prod_{t \in \cal{Q} \cap \cal{D}}
\hat{p}(t|M_D) \times \displaystyle \prod_{t \in \cal{Q} - \cal{D}}
\hat{p}(t|M_D) \\
 & = \prod_{t \in \cal{Q} \cap \cal{D}} \hat{p}(t|M_D) \times
\displaystyle \prod_{t \in \cal{Q} - \cal{D}} \frac{cf_t}{cs} \\
 & = \prod_{t \in \cal{Q} \cap \cal{D}} \hat{p}(t|M_D) \times
\displaystyle
     \frac {\prod_{t \in \cal{Q}} \frac{cf_t}{cs}} {\prod_{t \in \cal{Q} \cap
\cal{D}} \frac{cf_t}{cs}}
 \end{aligned}
\end{equation}
 The term  $\prod_{t \in \cal{Q}} \frac{cf_t}{cs}$ in the above formula is constant
for any specific query string, so it can be dropped, since the goal is to find 
most similar tuples by ranking them based on the similarity scores.
Therefore, equation (\ref{eqlm6}) can be written as follows:
 \begin{equation}
  \label{eqlm8}
\begin {aligned} 
\displaystyle
  \hat{p}(Q|M_D) & = \displaystyle \frac { \displaystyle \prod_{t \in \cal{Q}
\cap \cal{D}} \hat{p}(t|M_D) } {\displaystyle \prod_{t \in \cal{Q}
\cap \cal{D}} \frac{cf_t}{cs}}
 \times \frac{\displaystyle \prod_{\forall t \in \cal{D}}(1 - \hat{p}(t|M_D))}
{\displaystyle \prod_{t \in \cal{Q} \cap \cal{D}}(1 - \hat{p}(t|M_D))}
 \end{aligned}
\end{equation}

This transformation allows us to efficiently compute similar tuples
by just storing $\hat{p}(t | M_D )$ and $\frac{cf_t}{cs}$ for each pair of
$t$ and $D$. Thus, we create table \texttt{BASE\_PM(tid, token, pm, cfcs)} where
\texttt{pm} $ = \hat p(token \vert M_{tid})$ and \texttt{cfcs}
$=\frac{cf_{token}}{cs}$ as the final result of the preprocessing step. We also calculate and store the term $\prod_{\forall t \in \cal{D}}(1 - \hat p(t|M_D))$ during preprocessing in relation \texttt{BASE\_SUMCOMPBASE(tid, sumcompm)}.

The query-time SQL statement to calculate similarity scores is shown in Figure \ref{lm}. The subquery in the statement computes the three terms in equation \ref{eqlm8} that include intersection of query and tuple tokens and therefore needs a join between the two token tables. The fourth term in the equation is read from the table stored during the preprocessing as described above.
 
\begin{figure}
\centering
 \small \tt  
\begin{tabular}{|l|} \hline
INSERT INTO LM\_SCORES (tid, score)  \\
SELECT B1.tid2, EXP(B1.score + B2.sumcompm)  \\
FROM (SELECT P1.tid AS tid1, T2.tid AS tid2, \\
\hspace{0.65in}   SUM(LOG(P1.pm)) - SUM(LOG(1.0-P1.pm)) \\
\hspace{0.65in}   - SUM(LOG(P1.cfcs)) AS score  \\
\hspace{0.35in}      FROM BASE\_PM P1, QUERY\_TOKENS T2 \\
\hspace{0.35in}      WHERE P1.token = T2.token  \\
\hspace{0.35in}      GROUP BY P1.tid, T2.tid) B1, \\
\hspace{0.3in}     BASE\_SUMCOMPMBASE B2 \\
WHERE B1.tid1=B2.tid \\
\hline
\end{tabular}
\normalfont
\caption{SQL Code for Language Modeling}
\label{lm}
\end{figure}

\subsection{Hidden Markov Models}

We rewrite equation (\ref{eqhmm1}) as follows:
  \begin{equation} \small
  \begin{aligned}
 \label{eqhmm4}
 \displaystyle P(Q|D\text{ is similar}) & = \prod_{q \in \cal{Q}} {(a_0 P(q|GE)
+ a_1 P(q|D))} \\
 & = \displaystyle \prod_{q \in \cal{Q}} a_0 P(q|GE) \times \left[ \prod_{q \in
\cal{Q}} ({1 + \frac{a_1 P(q|D)}{a_0 P(q|GE)})} \right]
 \end{aligned}
 \end{equation}
 For a specific query, the term $\prod_{q \in \cal{Q}} a_0 P(q|GE)$ in the above
formula is constant for all
tuples in the base relation and therefore can be dropped since our goal is to
order tuples based on similarity to a specific query string. So the
modified similarity score will be:
\begin{eqnarray}
\nonumber  sim_{HMM}(Q,D) &=&  \prod_{q \in \cal{Q}} {(1 + \frac{a_1 P(q|D)}{a_0
P(q|GE)})}\\
\label{simhmm} &=&  \prod_{q \in \cal{Q} \cap \cal{D}} {(1 + \frac{a_1
P(q|D)}{a_0
P(q|GE)})}
\end{eqnarray}
 In Equation \ref{simhmm}, $q \in \cal{Q}$ changes to $q \in \cal{Q} \cap
\cal{D}$ because $P(q|D) = 0$ for all $q \notin \cal{D}$. Thus we can calculate
the term $(1 +
\frac{a_1 P(q|D)}{a_0 P(q|GE)})$ for all \texttt{tid, token} pairs during
preprocessing and store them as \texttt{weight} in relation
\texttt{BASE\_WEIGHTS(tid, token, weight)}. Notice that the term
$P(q|D)$ is equal to $\hat p_{ml}(q,D)$ in language modeling;
we use a relation \texttt{BASE\_PML(tid, token, pml)} for it. 
Calculating $P(q|GE)$ and storing it in relation \texttt{BASE\_PTGE(token, ptge)} is also fairly simple. The final SQL
query for preprocessing and the SQL statements for calculating similarity scores,
are shown in Figure \ref{hmm-SQL}.
 
\begin{figure}
\centering
 \small \tt  
\begin{tabular}{|ll|} \hline
\multicolumn{2}{|c|}{Preprocessing} \\ \hline
\multicolumn{2}{|l|}{INSERT INTO BASE\_WEIGHTS(tid,token,weight)}  \\
SELECT   & M2.tid, M2.token, \\
         & \hspace{0.04in} (1 + (a1*M2.pml) / (a0*P2.ptge))  \\
FROM     & BASE\_PTGE P2, BASE\_PML M2  \\
WHERE    & P2.token = M2.token  \\ \hline
\multicolumn{2}{|c|}{Query} \\ \hline
\multicolumn{2}{|l|}{INSERT INTO HMM\_SCORES (tid, score)} \\
SELECT   & W1.tid, T2.tid, EXP(SUM(LOG(W1.weight))) \\
FROM     & BASE\_WEIGHTS W1, QUERY\_TOKENS T2 \\
WHERE    & W1.token = T2.token \\
\multicolumn{2}{|l|}{GROUP BY  T2.tid, W1.tid} \\
\hline
\end{tabular}
\normalfont
\caption{SQL Code for HMM}
\label{hmm-SQL}
\end{figure}

\section{Edit-based Predicates}

We use the same declarative framework proposed in \cite{joinForFreeVLDB01} for
approximate matching based on edit-distance.
The idea is to use properties of q-grams created from the strings to generate a
candidate set in a way that no false negatives are guaranteed to exist
but the set may contain false positives. 
The set is subsequently filtered by computing the exact edit 
similarity score between the query and
the strings in the candidate set. Computing the edit similarity score
is performed using a UDF. The SQL statements for candidate set generation and
score calculation are available in \cite{joinForFreeVLDB01}.

\section{Combination Predicates}

Since the calculation of the score function for $GES$ (Equation \ref{GES}) between
a query string and all tuples in a relation could be very expensive, we can
first identify a candidate set of tuples similar to the methodology used for
edit-distance and then use a UDF to compute exact scores between the query string
and the strings in the candidate set. The elements of the candidate set are
selected using a threshold $\theta$ and the following score formula which
ignores the ordering between word tokens. This
formula over-estimates $sim_{GES}(Q,D)$ \cite{fmsSIGMOD03}:
\begin{equation} \small
 \label{eqges-jac}
 sim_{GES}^{Jaccard}(Q,D) = \frac{1}{wt(Q)} \sum_{t \in \cal{Q}} w(t) \cdot \max_{r
\in \cal{D}}
(\frac{2}{q}sim_{Jaccard}(t,r) + d_q )
\end{equation}
where $wt(Q)$ is the sum of weights of all {\em word} tokens in $Q$, 
$w(t)$ is the $idf$ weight for word token $t$, 
$q$ is a positive integer indicating the q-gram length
extracted from {\em words} in order to calculate $sim_{Jaccard}(t, r)$ 
and $d_q = (1 -1/q)$ is an adjustment term.
In order to enhance the performance of the operation, we can
employ min-wise independent permutations \cite{broder00minwise} 
to approximate $sim_{Jaccard}(t_1, t_2)$ in
Equation \ref{eqges-jac}. Description of min-wise independent permutations is beyond
the scope of this thesis. This would result in substituting $sim_{Jaccard}$ with
the min-hash similarity $sim_{mh}(t_1, t_2)$, which is a provable approximation.
The resulting metric, $GES^{apx}$, is shown to be an
upper-bound for $GES$ in expectation \cite{fmsSIGMOD03}:

\begin{equation}
 \label{eqges-apx}
 sim_{GES}^{apx}(Q,D) = \frac{1}{wt(Q)} \sum_{t \in \cal{Q}} w(t) \cdot \max_{r
\in \cal{D}}
(\frac{2}{q}sim_{mh}(t,r) + d_q )
\end{equation}

In order to implement the above predicates, we need to preprocess the relation
using the following methodology:
\begin{itemize}
\item Tokenization in two levels, first tokenizing into words and then
tokenizing each word into q-grams. Word tokens are stored in relation
\texttt{BASE\_TOKENS(tid, token)} and q-grams are stored in
\texttt{BASE\_QGRAMS(tid, token, qgram)}. 
\item Storing $idf$ weights of word tokens in relation
\texttt{BASE\_IDF (token,idf)} as well as the average of $idf$ weights in the base
relation to be used as $idf$ weights of unseen tokens.
 \item Calculating weights related to the similarity employed to compare 
tokens, i.e., $sim(t,r)$. For GES$^{Jaccard}$ employing the Jaccard predicate, 
this includes storing the number of
q-grams for each word token in relation \texttt{BASE\_TOKENSIZE (tid,
token, len)}. For GES$^{apx}$, we have to calculate 
minhash signatures (required by min-wise independent permutations). SQL statements 
for generating min-hash signatures and min-hash similarity scores, $sim_{mh}(t,r)$, are available in Appendix A.
\end{itemize}

We omit most of SQL statements inside this chapter. In order to make the presented
statements more readable, we assume that the following
auxiliary relations are available to us; in practice, they are calculated
on-the-fly as subqueries (refer to Appendix for complete
queries):

\begin{itemize}
 \item \texttt{QUERY\_IDF(token, idf)} stores $idf$ weights for each token in the
query. Weights are retrieved from the base weights relation and the average $idf$ 
value over all tokens in the base relation is used as the weight of query tokens not present in the base relation. \texttt{SUM\_IDF(token, sumidf)} will store sum of idf weights for query tokens.
\ignore{ 
\item \texttt{SIMILARITY\_SCORES(tid, token1, token2, sim)} stores a similarity
score of each token in the query with each token in each tuple in the base relation.
Such a score could have been computed using
the IntersectSize predicate, Jaccard predicate or its approximation using
min-wise independent permutations.
}

 \item \texttt{MAXSIM(tid, token, maxsim)} stores the maximum of the
similarity scores between the tokens in tuple \texttt{tid} and each \texttt{token} in the query.
\end{itemize}

The tables above do not have to be computed beforehand, they are rather computed on the fly
at query execution time. Assuming however they are available, 
the SQL statements for computing the scores for GES$^{apx}$, GES$^{Jaccard}$ are shown in Figure
\ref{gesquery}. 

\begin{figure}
\centering
 \small \tt  
\begin{tabular}{|ll|} \hline
\ignore{
\multicolumn{2}{|l|}{INSERT INTO MINHASH\_SCORES(tid1,token1,token2,sim)} \\
SELECT  & R1.TID, R1.TOKEN, R2.TOKEN, COUNT(*)/3  \\
FROM    & BASE\_MHASHSIGNATURE R1,  \\
        &  (SELECT Q.TID, Q.TOKEN, H.FID, \\
        &         \hspace{0.42in} MIN(H.VALUE) AS VALUE  \\
        &   FROM  \hspace{0.02in} QUERY\_QGRAMS Q, BASE\_HASHVALUE H  \\
        &   WHERE \hspace{0.02in} Q.QGRAM = H.QGRAM  \\
        &   GROUP BY \hspace{0.02in} Q.TID, Q.TOKEN, H.FID) R2  \\
WHERE   & R1.FID = R2.FID AND R1.VALUE = R2.VALUE  \\
\multicolumn{2}{|l|}{GROUP BY R1.TID, R1.TOKEN, R2.TID, R2.TOKEN} \\
 & \\
\multicolumn{2}{|l|}{INSERT INTO MAXSIM(tid1, token2, maxsim)} \\
SELECT & S.TID1, S.TOKEN2, MAX(SIM) AS maxsim  \\
FROM   & SIMILARITY\_SCORES S  \\
\multicolumn{2}{|l|}{GROUP BY S.TID1, S.TOKEN2, S.TID2} \\
 & \\}
 
\multicolumn{2}{|l|}{INSERT INTO GESAPX\_RESULTS(tid, score)}  \\
SELECT  & MS.tid, 1.0/SI.sumidf *\\
        & SUM(I.idf*(((2.0/$q$)*MS.maxsim)+(1-1/$q$))) \\
FROM    & MAXSIM MS, QUERY\_IDF I, SUM\_IDF SI \\
WHERE   & MS.token = I.token \\
\multicolumn{2}{|l|}{GROUP BY MS.tid, SI.sumidf} \\
\hline
\end{tabular}
\normalfont
\caption{SQL Code for GES$^{apx}$, GES$^{Jaccard}$}
\label{gesquery}
\end{figure}

 SoftTFIDF can also be implemented similar to GES approximation predicates. During preprocessing, we need to first tokenize the string into word tokens and store them in \texttt{BASE\_TOKENS(tid, token)}. Depending on the function used for similarity score  between word tokens, we may need to tokenize each word token into qgrams as well. We then need to store normalized tf-idf weights of tokens in the tuples in the base relation in \texttt{BASE\_WEIGHTS(tid, token, weight)}.

Here again, at query time, we assess the final score formula of equation (\ref{softtfidf}), in a single SQL statement. For presentation purposes, assume that the following relations have been materialized:
\begin{itemize}
 \item \texttt{QUERY\_WEIGHTS(token, weight)} stores normalized tf-idf weights for each \texttt{token} in the query table.
 \item \texttt{CLOSE\_SIM\_SCORES(tid, token1, token2, sim)} stores the similarity
score of each token in the query (\texttt{token2}) with each token of each tuple in the base relation, 
where the score is greater than a threshold $\theta$ ($\theta$ specified at query time).
Such a score could have been computed using a declarative realization of some
other similarity predicate or a 
UDF to compute similarity using a string distance scheme (e.g., Jaro-Winkler \cite{Winkler99}).
 \item \texttt{MAXSIM(tid, token, maxsim)} stores the maximum of the \texttt{sim} score for each 
query \texttt{token} among all \texttt{tid}s in \texttt{CLOSE\_SIM\_SCORES} relation. \texttt{MAXTOKEN(tid, token1, token2, maxsim)} stores $\arg\max_{r \in \cal{\texttt{tid}}} (sim(token2,r))$ as well, i.e., the token in each tuple in the base relation that has the maximum similarity with a query token \texttt{token2} in $CLOSE(\theta,Q,D)$
\end{itemize}

Figure \ref{softtfidfquery} shows the SQL statement for \texttt{MAXTOKEN} table and the final similarity score for SoftTFIDF.

\begin{figure}
\centering
 \small \tt  
\begin{tabular}{|ll|} \hline
\ignore{
\multicolumn{2}{|l|}{INSERT INTO MAXSIM(tid, token2, maxsim)} \\
SELECT & S.TID, S.TOKEN2, MAX(SIM) AS maxsim  \\
FROM   & CLOSE\_SIM\_SCORES S  \\
\multicolumn{2}{|l|}{GROUP BY S.TID, S.TOKEN2, S.TID2} \\
 & \\
 }
 \multicolumn{2}{|l|}{INSERT INTO MAXTOKEN(tid,token1,token2,maxsim)} \\
SELECT & CS.tid, CS.token1, \\ 
       & CS.token2, MS.maxsim \\
FROM   & MAXSIM MS, CLOSE\_SIM\_SCORES CS \\
WHERE  & CS.tid=MS.tid AND \\
       & CS.token2=MS.token2 AND MS.maxsim=CS.sim \\
 & \\

\multicolumn{2}{|l|}{INSERT INTO SoftTFIDF\_RESULTS (tid, score)} \\
SELECT   & TM.tid, SUM(I.weight*WB.weight*TM.maxsim) \\
FROM     & MAXTOKEN TM, \\
         & QUERY\_WEIGHTS I, BASE\_WEIGHTS WB \\
WHERE    & TM.token2 = I.token AND TM.tid = WB.tid \\
         & AND TM.token1 = WB.token \\
\multicolumn{2}{|l|}{GROUP BY TM.tid} \\

\hline
\end{tabular}
\normalfont
\caption{SQL Code for SoftTFIDF - Query time }
\label{softtfidfquery}
\end{figure}

\chapter{Evaluation}
\label{experimental}

We experimentally evaluate the performance of each of the similarity
predicates presented thus far and compare their accuracy. The choice of the best
similarity predicate in terms of accuracy highly depends on the
type of datasets and errors present in them. The choice in terms of performance
depends on the characteristics of specific predicates.
We therefore evaluate the (a) accuracy of predicates using different datasets with
different error characteristics and the (b) performance by dividing the preprocessing and
query execution time into various phases to obtain detailed understanding
on the relative benefits and limitations. All our experiments are performed
on a desktop PC running MySQL server 5.0.16 database system over Windows XP SP2
with Pentium D 3.2GHz CPU and 2GBs of RAM.

\section{Benchmark}

In the absence of a common benchmark for data cleaning, we resort
to the definition of our own data generation scheme with controlled error.
In order to generate datasets for our experiments, we modify and significantly
enhance the UIS database generator which has effectively been used in the past to evaluate duplicate detection algorithms
\cite{MergePurge}. We use the data generator to inject different types and
percentages of errors to a clean database of string attributes. We keep track of the source tuple from which the
erroneous tuples have been generated in order to determine precision and recall
required to quantify the accuracy of different predicates. The generator allows to create data sets of varying sizes and error types, thus
is a very flexible tool for our evaluation. The data generator accepts clean tuples and generates erroneous duplicates based
on a set of parameters. Our data generator provides the following parameters to control the error injected in the data:
\begin{itemize}
\item {the size of the dataset to be generated}
\item {the fraction of clean tuples to be utilized to generate erroneous duplicates}
\item \textit{distribution of duplicates}: the number of duplicates generated
for a clean tuple can follow a uniform, Zipfian or Poisson distribution.
\item \textit{percentage of erroneous duplicates}: the fraction of duplicate
tuples in which errors are injected by the data generator.
\item \textit{extent of error in each erroneous tuple}: the percentage of characters
that will be selected for injecting character edit error (character insertion,
deletion, replacement or swap) in each tuple selected for error injection.
\item \textit{token swap error}: the percentage of word pairs that will be
swapped in each tuple that is selected for error injection.
\end{itemize}

We use two different sources of data: a data set
consisting of \textit{company names} and a data set consisting of
\textit{DBLP Titles}. Statistical details for the two datasets are
shown in Table \ref{statTable}. For the company names dataset, we also inject domain specific \textit{abbreviation
errors}, e.g., replacing \texttt{Inc.} with \texttt{Incorporated} and vice
versa.

\begin{table}
\small
\centering
\begin{tabular}{|c|c|c|c|} \hline
 dataset & \#tuples & Avg. tuple length & \#words/tuple \\ \hline
 Company Names & 2139 & 21.03 & 2.92 \\ \hline
 DBLP Titles & 10425 & 33.55 & 4.53\\ \hline
\end{tabular}
\caption{Statistics of Clean Datasets} 
\label{statTable}

\end{table}

For both datasets, we generate different erroneous datasets by varying the
parameters of the data generator as shown in Table \ref{paramS}.
\begin{table}
\begin{center}
\small
\begin{tabular}{|c|c|}  \hline
parameter & range \\ \hline
size of dataset & 5k - 100k \\ \hline
\# clean tuples & 500 - 10000 \\ \hline
duplicate distribution & uniform, Zipfian \\ \hline
erroneous duplicates & 10\% - 90\% \\ \hline
extent of error per tuple & 5\% - 30\% \\ \hline
token swap error & 10\% - 50\% \\ \hline
\end{tabular}
\caption{Range of Parameters Used For Erroneous Datasets} 
\label{paramS}
\end{center}
\end{table}

We show accuracy results for 8 different erroneous
datasets generated from a data set of company names, each containing 5000 tuples
generated from 500 clean records, with uniform distribution.
We choose to limit the size of the data sets 
to facilitate experiments and data collection since each
experiment is run multiple times to obtain statistical
significance. We conducted experiments with data sets of
increasing size and we observed that the overall accuracy trend
presented remains the same. We consider the results presented highly
representative across erroneous data sets (generated according to
our methodology) of varying sizes, and duplicate distributions.
We classify these 8 datasets into  \emph{dirty}, \emph{medium} and
\emph{low} error datasets
based on the parameters of data generation. We have also generated 5 datasets,
each having only one specific type of error, in order to evaluate the effect of
specific error types. Table \ref{classTable} provides more details on the
datasets. Table \ref{sampleTable} shows a sample
of duplicates generated by the data generator from $CU1$ and $CU5$.

\begin{table}

\centering \small
\begin{tabular}{|c|c|c|c|c|c|} \hline
  &           & \multicolumn{4}{|c|} {Percentage of} \\
\cline{3-6}
 Class & Name & erroneous  & errors in & token & Abbr. \\
     &       & duplicates   & duplicates & swap & error \\ \hline
 Dirty &  CU1 & 90 & 30 & 20 & 50 \\ \hline
 Dirty &  CU2 & 50 & 30 & 20 & 50 \\ \hline
 Medium &  CU3 & 30 & 30 & 20 & 50 \\ \hline
 Medium &  CU4 & 10 & 30 & 20 & 50 \\ \hline
 Medium &  CU5 & 90 & 10 & 20 & 50 \\ \hline
 Medium &  CU6 & 50 & 10 & 20 & 50 \\ \hline
 Low  &  CU7 & 30 & 10 & 20 & 50 \\ \hline
 Low  &  CU8 & 10 & 10 & 20 & 50 \\ \hline
 -  &  F1 & 50 & 0 & 0 & 50 \\ \hline
 -  &  F2 & 50 & 0 & 20 & 0 \\ \hline
 -  &  F3 & 50 & 10 & 0 & 0 \\ \hline
 -  &  F4 & 50 & 20 & 0 & 0 \\ \hline
 -  &  F5 & 50 & 30 & 0 & 0 \\ \hline
\end{tabular}
\caption{Classification of Datasets} 
\label{classTable}

\end{table}

\begin{table}
\centering 
\small
\begin{tabular}{|c|c|} \hline \texttt 
 & CU1 \\ \hline
$t_1^1$ & Stsalney Morgan cncorporsated Group \\
$t_2^1$ & jMorgank Stanlwey Grouio Inc.     \\
 $t_3^1$ & Morgan Stanley Group Inc.           \\
$t_4^1$ & Sanlne Morganj Inocrorpated Group \\
 $t_5^1$ & Sgalet Morgan Icnorporated Group  \\ \hline
 & CU5 \\ \hline
$t_1^5$ & Morgan Stanle Grop Incorporated  \\
$t_2^5$ &  Stalney Morgan Group Inc. \\
 $t_3^5$ & Morgan Stanley Group In. \\
 $t_4^5$ & Stanley Moragn Grou Inc. \\
 $t_5^5$ & Morgan Stanley Group Inc. \\ \hline
\end{tabular}

\caption{Sample Tuples from CU1 \& CU5 Datasets}
\label{sampleTable}

\end{table}

\section{Evaluating Accuracy}

We measure the accuracy of predicates, utilizing known methods from the
information retrieval literature in accordance to common practice in IR
\cite{irbook}. We compute the \emph{Mean Average Precision (MAP)} and \emph{Mean Maximum F$_1$} scores of the
rankings of each dataset imposed by approximate selection queries utilizing
our predicates. Average Precision(AP), is the average of the precision after each similar record is retrieved, i.e.,
\begin{equation}
\displaystyle \frac{\sum_{r=1}^{N}[P(r) \times rel(r)]}{\textnormal{number of
relevant records}}
\end{equation}
 where $N$ is the total number of records returned, $r$ is the rank of the
record, i.e., the position of the record in the result list sorted by
decreasing similarity score, $P(r)$ is the precision at rank $r$, i.e.,
the ratio of the number of \emph{relevant} records having rank $\le r$ to the
\emph{total} number of records having rank $\le r$, and $rel(r)$ is 1 if the
record at rank $r$ is relevant to the query and 0 otherwise. This measure
emphasizes returning more similar strings earlier. MAP is the mean AP value over a set of queries. 
Maximum F$_1$ measure is the maximum F$_1$ score (the harmonic mean of precision and recall)
over the ranking of records, i.e.,
\begin{equation}
 \displaystyle \max_r{ [ \frac{2 \times Pr(r) \times Re(r)}{ Pr(r) + Re(r)} ] }
\end{equation}
where $Pr(r)$ and $Re(r)$ are precision and recall values for rank
$r$. $Pr(r)$ is as defined above. $Re(r)$ is the ratio of the number of relevant
records having rank $\le r$ to the total number of relevant records. Again, we
compute mean maximum F$_1$ over a set of queries.

Our data generation methodology allows to associate easily a clean
tuple with all erroneous versions of the tuple generated using
our data generator.A clean tuple and its erroneous duplicates are assigned
the same cluster id. Essentially each time we pick a tuple from a
cluster, using its string attribute as a query we consider all the
tuples in the same cluster (tuples with the same cluster id) as relevant to this query.
For each query and a specific predicate, we return a list of tuples sorted in the order of
decreasing similarity scores.
Thus, it is easy to identify relevant and irrelevant records among the 
results returned for a specific query and similarity predicate.
In order to maintain our evaluation independent of any threshold constants
(specified in approximate selection predicates) we do not prune this list utilizing thresholds.
For each dataset, we compute the mean average precision and mean
maximum F$_1$ measure over 500 randomly selected queries taken from that data
set (notice that our query workload contains
both clean as well as erroneous tuples).  Thus, our accuracy results
represent the {\em expected} behaviour of the predicates over queries
and thresholds.
We report the values for MAP only since the results were consistently
similar for max F1 measure in all our experiments.

\section{Settings}

\subsection{Choice of Weights for Weighted Overlap Predicates}

Both WeightedMatch (\texttt{WM}) and
WeightedJaccard (\texttt{WJ}) predicates
require a weighting scheme to assign weights to the tokens. It is desirable to
use a weighting scheme which captures the importance of tokens. We experimented with $idf$
and the Robertson-Spark Jones ($RS$) weighting scheme given in Equation \ref{rs_wt}
 and found that $RS$ weights lead to better accuracy. So in the following
discussion, we use $RS$ weights for weighted overlap predicates.

\subsection{Parameter Settings for Predicates}

\ignore{Several predicates presented in Chapter \ref{framework} depend on
various independent parameters. Though previous works describing these predicates give some
heuristics to choose these parameters, we experimentally determined these
parameters by varying  them in order to get the best performance on our data.
The values we obtained are close to the
values proposed by the previous works.}
For all predicates proposed previously in the literature we set any  parameter values they
require for tuning as suggested in the respective papers. For the predicates
presented herein for data cleaning tasks, for the case of
BM25, we set $k_1$=1.5,  $k_3$=8 and $b$=0.675; for HMM, we set $a_0$ to
0.2, although our experiments  show that the accuracy results are not very
sensitive to the value of $a_0$ as long as a reasonable value is chosen (i.e., a
value not close to 0 or 1).

The SoftTFIDF predicate requires a similarity predicate over the word tokens. We
experimented with various similarity predicates like Jaccard, IntersectSize, edit distance,
Jaro-Winkler, etc. and choose Jaro-Winkler since SoftTFIDF with Jaro-Winkler
(\texttt{STfIdf w/JW}) performs the best. This was also observed  in \cite{cohen1}.
Two words are similar in SoftTFIDF if their similarity score exceed a given
threshold $\theta$. SoftTFIDF with Jaro-Winkler
performed the best with $\theta$=0.8.  Finally, we set $c_{ins}$ for GES predicate to 0.5 as
proposed in \cite{fmsSIGMOD03}. For calculating accuracy, we use the exact GES as shown in Equation
\ref{GES}. We remark that we do not prune the results based on any threshold
in order to keep the evaluation independent of the threshold values.

\subsection{Q-gram Generation}

\label{qgramgen}
Qgram generation  is a common preprocessing step for all
predicates. We use an SQL statement  similar to that presented in
\cite{joinForFreeVLDB01} to generate q-grams, with a slightly different
approach. We first insert $q-1$  special symbols (e.g. $\$$) in
place of all whitespaces in each string, as well as at the beginning and end of the
strings. In this way we can fully capture all errors caused by different orders
of words, e.g., ``Department of Computer Science" and ``Computer Science Department". 
For qgram generation we also need to have an optimal value of
qgram size ($q$). A lower value of $q$ ignores the ordering
of characters in the string while a higher value can not capture
the edit errors. So an optimum value is required to capture
the edit errors taking in account the ordering of
characters in the string. The table below shows the accuracy
comparison of different qgram based predicates  (Jaccard, tf-idf
(\texttt{Cosine}), HMM and BM25) in the dirty cluster of our data
sets:
\begin{center}
\small
\begin{tabular}{|c|c|c|c|c|c|} \hline
$q$ & Jaccard & Cosine & HMM & BM25 \\ \hline
2 & 0.736 & 0.783 & 0.835 & 0.840  \\ \hline
3 & 0.671 & 0.769 & 0.807 & 0.805 \\ \hline 
\end{tabular}
\end{center}

The trend is similar for other predicates and the accuracy further
drops for higher values of $q$. Thus, we set $q$=2 as
it achieves the best accuracy results.

\begin{table*}
\centering \small
\begin{tabular}{|c|c|c|c|c|c|c|c|c|c|c|}  \hline
Type of Error &  Xect & Jac. & WM & WJ & Cosine, BM25, & ED & GES & S
TfIdf \\ 
 &  &  &  & & LM, HMM &  & & w/JW \\ \hline
abbr. error (F1) & 0.94 & 0.96 & 0.98 & 1.0 & 1.0 & 0.89 & 1.0 & 1.0\\ \hline
token swap error (F2) &  1.0 & 1.0 & 1.0 & 1.0 & 1.0 & 0.77 & 0.94 & 1.0\\
\hline
\end{tabular}
\caption{Accuracy: Abbr. and Token Swap Errors \label{tababbts}}
\end{table*}

\section{Accuracy Results}

In this section we present a detailed comparison of the
effectiveness of the similarity predicates in capturing
the different types of error introduced in the data.
\ignore{Table \ref{anacccmp} presents an intuitive comparison of the types
of error captured by different similarity predicates.}

\begin{figure*}[!th]
\centering
\begin{tabular}{c}
\epsfig{file=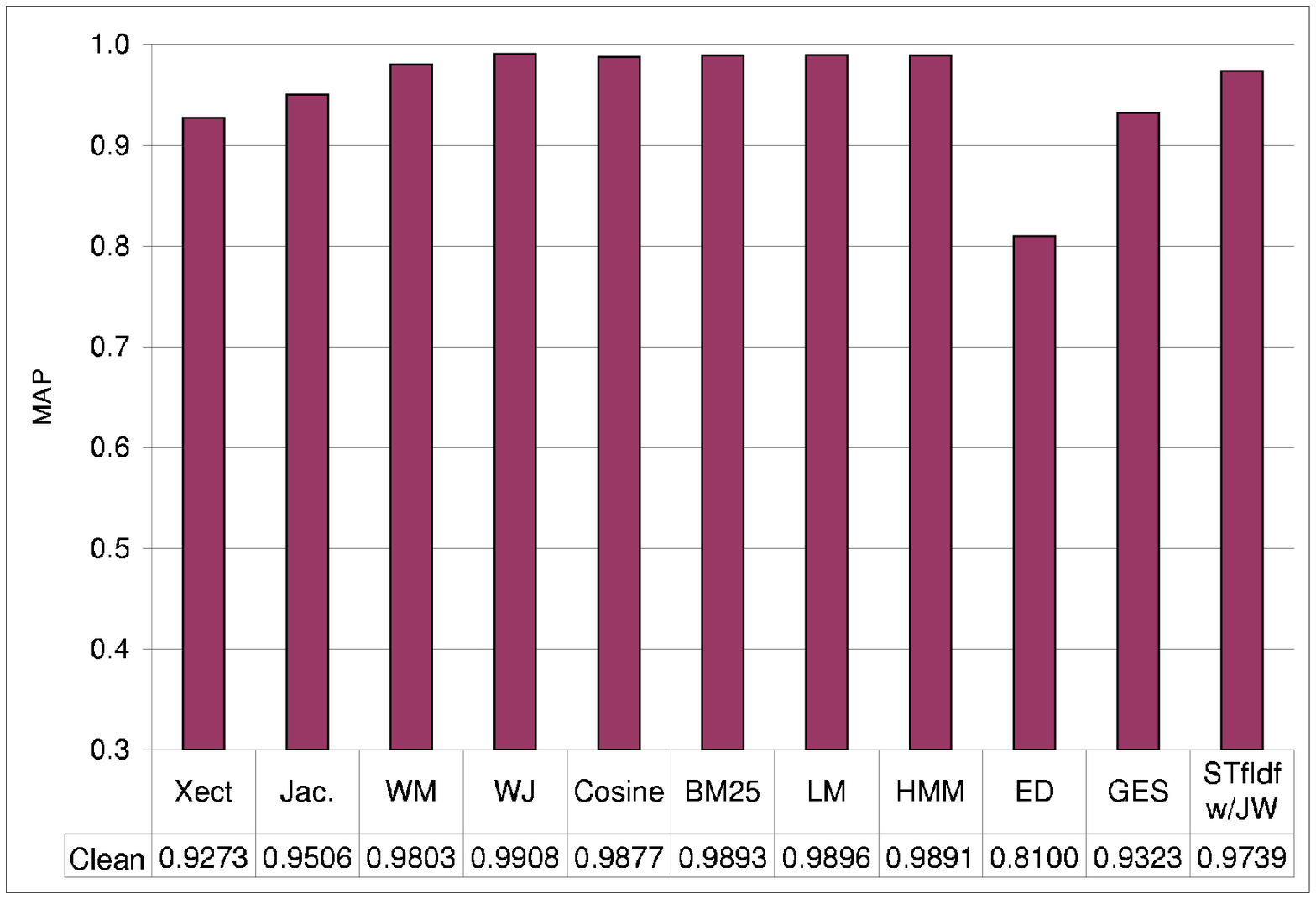,width=3.5in} \\
 (a) Low Error Datasets \\
\epsfig{file=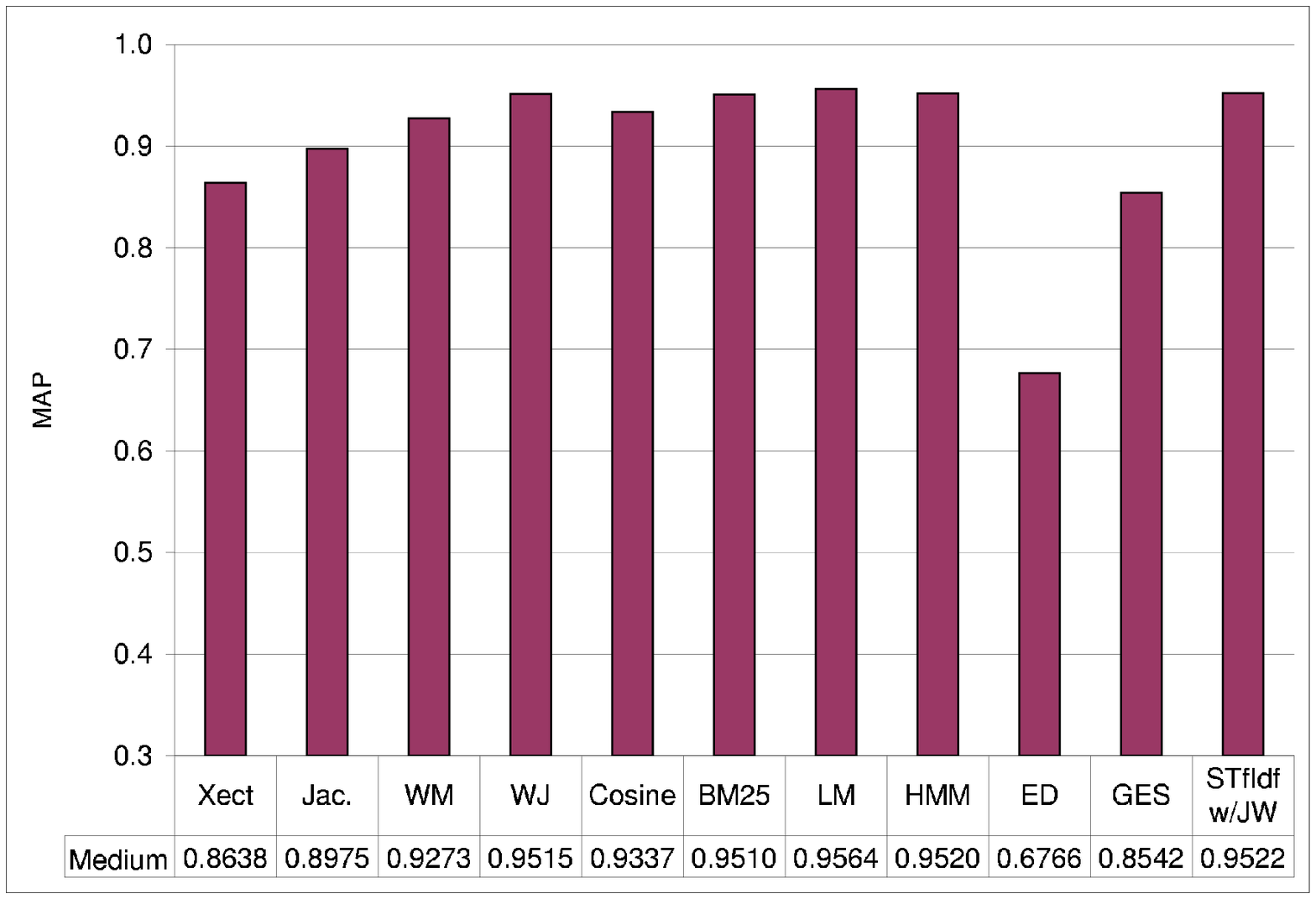,width=3.5in} \\
 (b) Medium Error Datasets \\
\epsfig{file=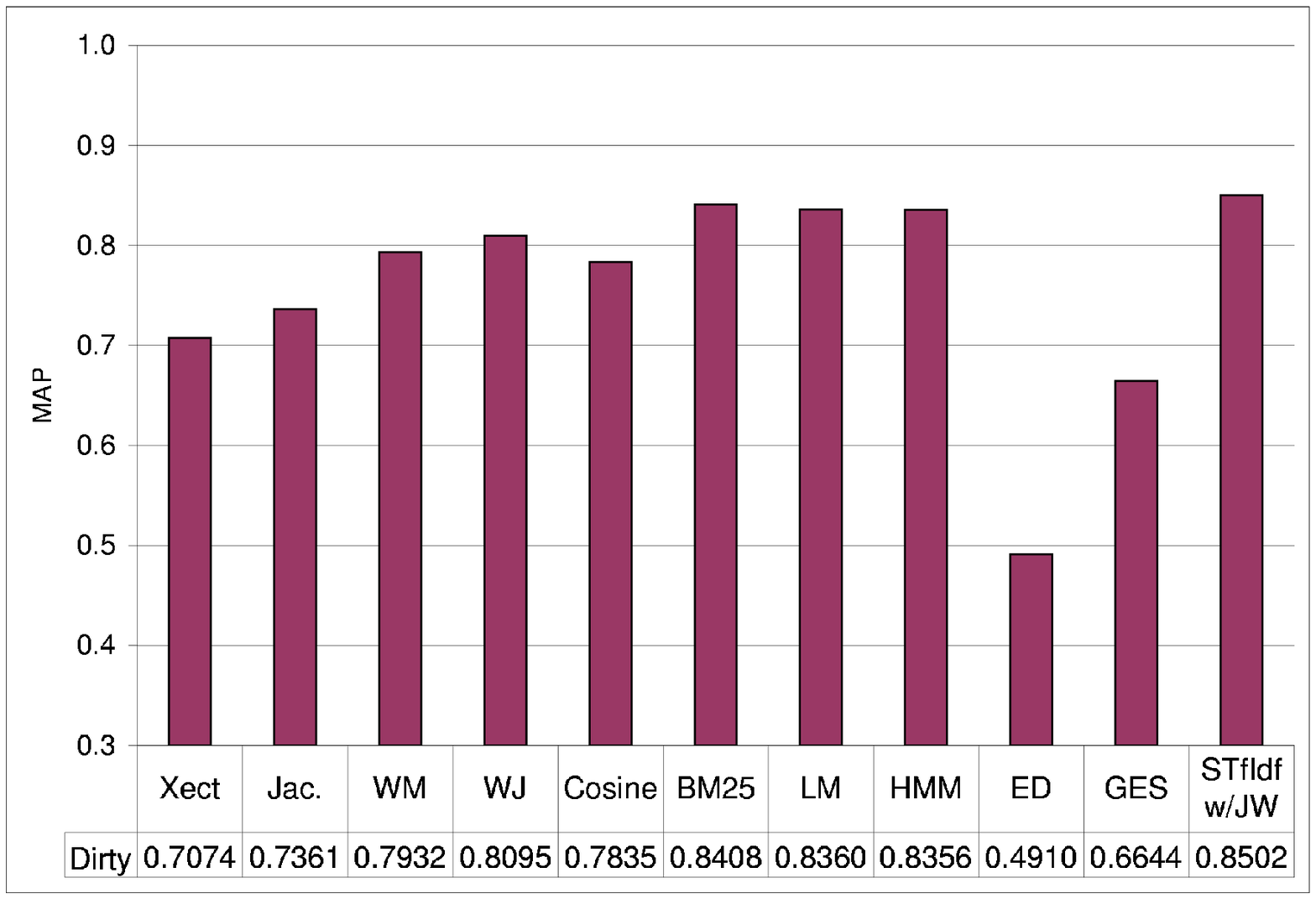,width=3.5in} \\
 (c) Dirty Datasets \\
\end{tabular}
\caption{MAP values for different predicates on different datasets}
\label{accchart1}

\end{figure*}

\textbf{Abbreviation error}: Due to abbreviation errors, a tuple
\texttt{AT\&T}  \texttt{Incorporated} gets converted to \texttt{AT\&T Inc}. Note
that\\ \texttt{Incorporated} and \texttt{Inc} are frequent words in the company
names database. For the query \texttt{AT\&T} \texttt{Incorporated}, the unweighted
overlap predicates Jaccard (\texttt{Jac.}) and IntersectSize (\texttt{Xect}) will assign to the tuple
\texttt{IBM} \texttt{Incorporated} greater similarity score than to the tuple \texttt{AT\&T Inc} since they just try to
match tuples on the basis of common qgrams. Edit distance (\texttt{ED}) will behave
similarly since it is cheaper to convert \texttt{AT\&T} \texttt{Incorporated} to \texttt{IBM}
\texttt{Incorporated} than to \texttt{AT\&T Inc}. The weight based predicates  are robust to 
abbreviation errors since they assign high weights to tokens
corresponding to rare (important) words e.g. \texttt{AT\&T}.  Table
\ref{tababbts} presents the accuracy of the predicates for the case of a data set
with only abbreviation error (dataset \texttt{F1}). All other predicates WeightedMatch(\texttt{WM}),
WeightedJaccrd(\texttt{WJ}), tf-idf(\texttt{Cosine}), BM25, HMM,
Language Modeling(\texttt{LM}) and SoftTFIDF(\texttt{STfIdf w/JW}) had near
perfect accuracy. Similar behaviour is observed when the percentage of duplicates and abbreviation
error is varied.

\textbf{Token swap errors}: Due to token swap errors, a tuple \texttt{Beijing
Hotel} gets converted to \texttt{Hotel Beijing}. Suppose there is a tuple
\texttt{Beijing Labs} present in the database, where \texttt{Labs} and
\texttt{Hotel} are equally important tokens but more
frequent than \texttt{Beijing}. For a query \texttt{Beijing Hotel},
edit distance and GES will
claim \texttt{Beijing Labs} more similar to the query than \texttt{Hotel
Beijing}. We remark that for accuracy calculation, we use exact GES as shown in
Equation \ref{GES}. All other predicates ignore the order of words, and hence will perform well for
token swap errors. Table \ref{tababbts}
shows the accuracy of the predicates for a data set with only token swap errors
(dataset \texttt{F2}). All other predicates had near perfect accuracy. Similar
trend is observed when the percentage of duplicates and token swap
error is varied.

\begin{table}
\centering \small
\begin{tabular}{|c|c|c|c|}  \hline
Predicate group & F3 & F4 & F5 \\ \hline
GES & 1.0 & .99 & .97 \\ \hline
BM25, HMM, LM, STfIdf w/JW & 1.0 & .97 & .91 \\ \hline
edit distance & .99 & .97 & .90 \\ \hline
WM , WJ, Cosine & .99 & .93 & .85 \\ \hline
Jaccard (\texttt{Jac.}), IntersectSize (\texttt{Xect}) & .99 & .91 & .81 \\ \hline
\end{tabular}
\caption{Accuracy: Only Edit Errors 
\label{tabedit}}
\end{table}

\textbf{Edit errors}: Edit errors involve character
insertion/ deletion/ replacement and character swap. The number of positions of
a string at which edit error has occurred defines the extent of the edit error.
All the predicates discussed above are robust towards low edit errors but
their accuracy degrades as the extent of edit error increases. Table \ref{tabedit} shows the
accuracy result for various predicates for increasing edit error in the data 
(datasets  \texttt{F3},  \texttt{F4} and  \texttt{F5}). The predicates giving near equal accuracy are grouped
together. GES is most resilient to edit errors. Edit distance, designed
to capture edit errors has average performance. BM25, STfIdf w/JW, and
probabilistic predicates (\texttt{LM} and \texttt{HMM}) are competitive in catching edit
errors and perform slightly better than edit distance.
The weighted overlap predicates (\texttt{WM} and \texttt{WJ}) with
$RS$ weights perform equivalent to tf-idf (Cosine) but not as good as edit
distance. Finally the unweighted overlap predicates Jaccard and IntersectSize perform the
worst as they ignore the importance of tokens. Similar
trend is observed when the percentage of erroneous duplicate is varied.

\subsection{Comparison of predicates}

Figure \ref{accchart1} shows MAP values for different predicates for the 3
classes of erroneous datasets described in Table \ref{classTable}. For the low error
datasets, all the predicates perform well except edit distance, GES, IntersectSize
and Jaccard. GES performs a little worse due to the presence of token swap
errors, IntersectSize and Jaccard perform worse because of abbreviation errors and
edit distance is the worst because of both factors.

When the error increases, the three types of errors occur in combination and
edit based predicates experience large accuracy degradation. 
The edit based predicates are already not good
at handling token swap errors and the presence of edit errors deteriorates their
effectiveness since the word token weights are no longer valid. This is not the
case for the qgram based predicates since edit errors affect only a small
fraction of qgrams and the remaining qgram weights are still valid. Consider a
query
$Q$=\texttt{Morgan Stanley Group Inc.} over dataset $CU5$, where we expect to
fetch the tuples shown in Table \ref{sampleTable}. The qgram based predicates are
able to return all the tuples at the top 5 positions in the list
according to similarity values. GES is not able to capture the token swap and
it ranks $t_2^5$ and $t_4^5$ at position 27 and 28 respectively. The edit
distance predicate performs worse; both $t_2^5$ and $t_4^5$ are absent from
the list of top 40 similar tuples. Both edit based predicates give high
similarity score to tuples like \texttt{Silicon Valley Group, Inc.} for
query $Q$ primarily because of low edit distance between \texttt{Stanley} and
\texttt{Valley}.

\ignore{Consider the
presence of two clean tuples in the database: $t_1$\texttt{Discovery
Comminications, Inc.} and $t_2$=\textt{TCI Communications, Inc.} Now due to the
combined error, $t_3$=\texttt{Discovery Cojumnicaions, Incorporated} and
$t_4$=\texttt{Coummnications, Discovery Inc.} are also present in the base
relation. In $t_3$ all the errors are present but $t_4$ does not have any
abbreviation error.  For a query $Q$=\texttt{Discovery Comminications, Inc.},
both edit distance and GES will rank $t_2$ more similar to $Q$ than $t_4$,
}

The unweighted overlap predicates ignore the importance of qgrams and hence
perform worse than the predicates that incorporate weights.  
It is interesting to note that the weighted
overlap predicates perform better than the tf-idf (\texttt{cosine}) predicate.
This is due to the $RS$ weighting scheme (Equation \ref{rs_wt}) for weight assignment of tokens
which has been shown more accurate than the $idf$ weighting scheme. 
The former captures importance of tokens more accurately than the
latter. The language modeling predicates (HMM and LM), and BM25 are 
always the best in all the three datasets.
The success of the SoftTFIDF is attributed to the underlying
Jaro-Winkler word level similarity predicate which can match the words
accurately even in the presence of high errors.

We also experimented with the GES$^{Jaccard}$ and GES$^{apx}$. Both
predicates make use of a threshold $\theta$ to prune irrelevant records
without calculating the exact scores. Depending on the value of $\theta$,
relevant records might also be pruned leading to a drop in accuracy.
Table \ref{GESthreshold} shows the variation in accuracy for  GES$^{Jaccard}$ and GES$^{apx}$
for threshold values ($\theta$) 0.7, 0.8 and 0.9 for dataset $CU1$ for which
GES (with no threshold) has 69.7\% accuracy. For GES$^{apx}$ we used 5 min hash
signatures in order to approximate the GES$^{Jaccard}$. We observe that increasing the number of
min-hash signatures takes more time without having a significant
impact on accuracy (pretty soon it demonstrates diminishing returns).
A small number of min hash signatures results in significant accuracy loss.

\begin{table}
\small
\begin{center}

\begin{tabular}{|c|c|c|c|} \hline
Predicate & $\theta=$0.7 & $\theta=$0.8 & $\theta=$0.9 \\  \hline
GES$^{Jaccard}$ & 0.692 & 0.683 & 0.603 \\ \hline
GES$^{apx}$ & 0.678 & 0.665 & 0.608 \\ \hline
\end{tabular}
\caption{Accuracy of GES Predicates for Different Thresholds}
\label{GESthreshold}
\end{center}
\end{table}

Experimental results show that for suitable thresholds GES$^{Jaccard}$ performs
as good as GES and the accuracy drops as the threshold increases. GES$^{apx}$,
being an approximation for GES$^{Jaccard}$, performs slightly worse than
GES$^{Jaccard}$. Similar results were observed for other datasets.

\section{Performance Results}

In this section, we compare different predicates based on preprocessing
time, query time and how well they scale when the size of the base table grows. As
expected, the performance depends primarily on the size of the
base table. Performance observations and trends remain relatively independent from
the error rate of the underlying data sets. Thus, we present the experiments on the DBLP datasets
with increasing size and medium amount of errors: 70\% of erroneous
duplicates, 20\% extent of error, 20\%  token swap error and no abbreviation
error.

\subsection{Preprocessing} 

We divide preprocessing time for a data set to make it amenable for
approximate selection queries into two phases. In the first phase,
tokenization is performed. Qgrams are extracted from strings in the way described
in section \ref{qgramgen} and stored in related tables. Aggregate weighted
(Cosine and  BM25) and language modeling predicates (LM and HMM) are fastest in this phase,
followed by overlap predicates (Xect and Jac.) with a small difference which is due to storing
distinct tokens only.
Combination predicates (\texttt{GES Jac}, \texttt{GES apx} and \texttt{STfIdf
w/JW}) are considerably slower in this phase since they involve
an extra level of tokenization into words. 

\begin{figure}

\centering
\epsfig{file=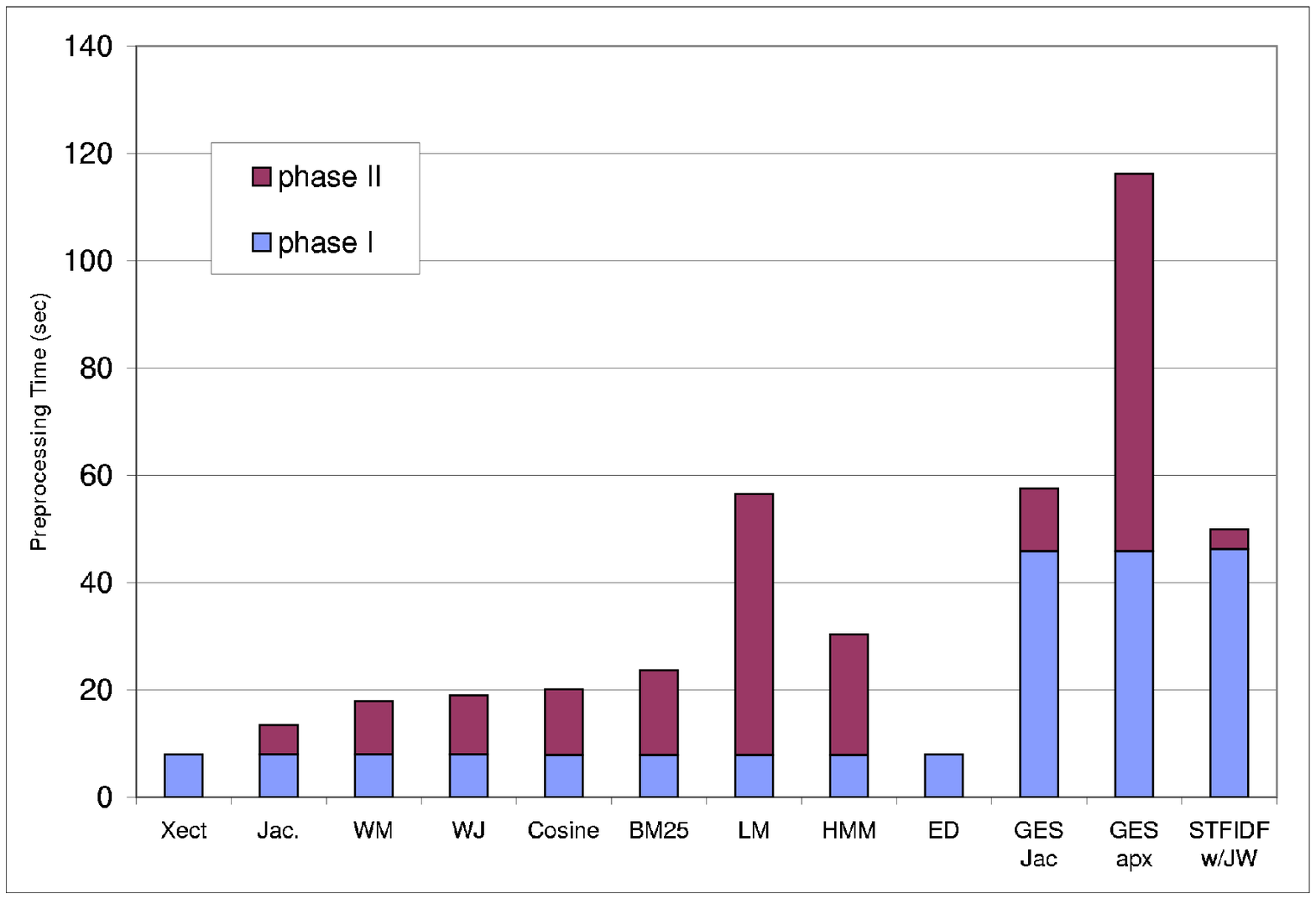, width=3.5in}
\caption{\small{Preprocessing time of different predicates}}
\label{prepchart}

\end{figure}

In the next phase, related weights are
calculated and assigned to tokens. In this phase, the fastest predicates are the overlap
predicates and edit distance (ED) followed by $GES^{Jaccard}$ and SoftTFIDF that
only require weight calculation for word tokens. Aggregate weighted and language
modeling predicates are considerably slower since calculating weights in these
predicates involves a lot of computation and creation of
many intermediate tables. Language modeling (LM) is the slowest  predicate among
probabilistic predicates since it requires the maximum number of intermediate tables to be created
and stored. $GES^{apx}$ requires to compute min-hash signatures for the
tokens separately for a number of hash functions on top of the two level
tokenization and IDF weight calculation, so it is the slowest of all predicates.
Figure \ref{prepchart} shows the preprocessing times for all predicates on a dataset
of 10,000 records with an average length of 37 characters. $GES^{apx}$ in
this Figure employs min-hash computation utilizing 5 hash functions (min hash signature size of 5). 
Preprocessing time for $GES^{apx}$ increases with increasing number of hash functions employed
for min-hash signature calculation.

\subsection{Query time}

\label{qanalysis}

Query time for a predicate is the time taken to rank the tuples from the base
table according to decreasing similarity score. Query time can also be
divided into two phases: preprocessing the query string
and computing similarity scores. The preprocessing part can itself be divided
into tokenization and weights computation phases as done for preprocessing
of the base relation. We didn't experience large variability in the time for 
query preprocessing among all predicates. As described in section \ref{lmdeclarative} the score formulas
for Language modeling and HMM are suitably modified by dropping query dependent terms which do not
alter the similarity score order and hence, the accuracy of the predicates.
 
\begin{figure}

\centering
\epsfig{file=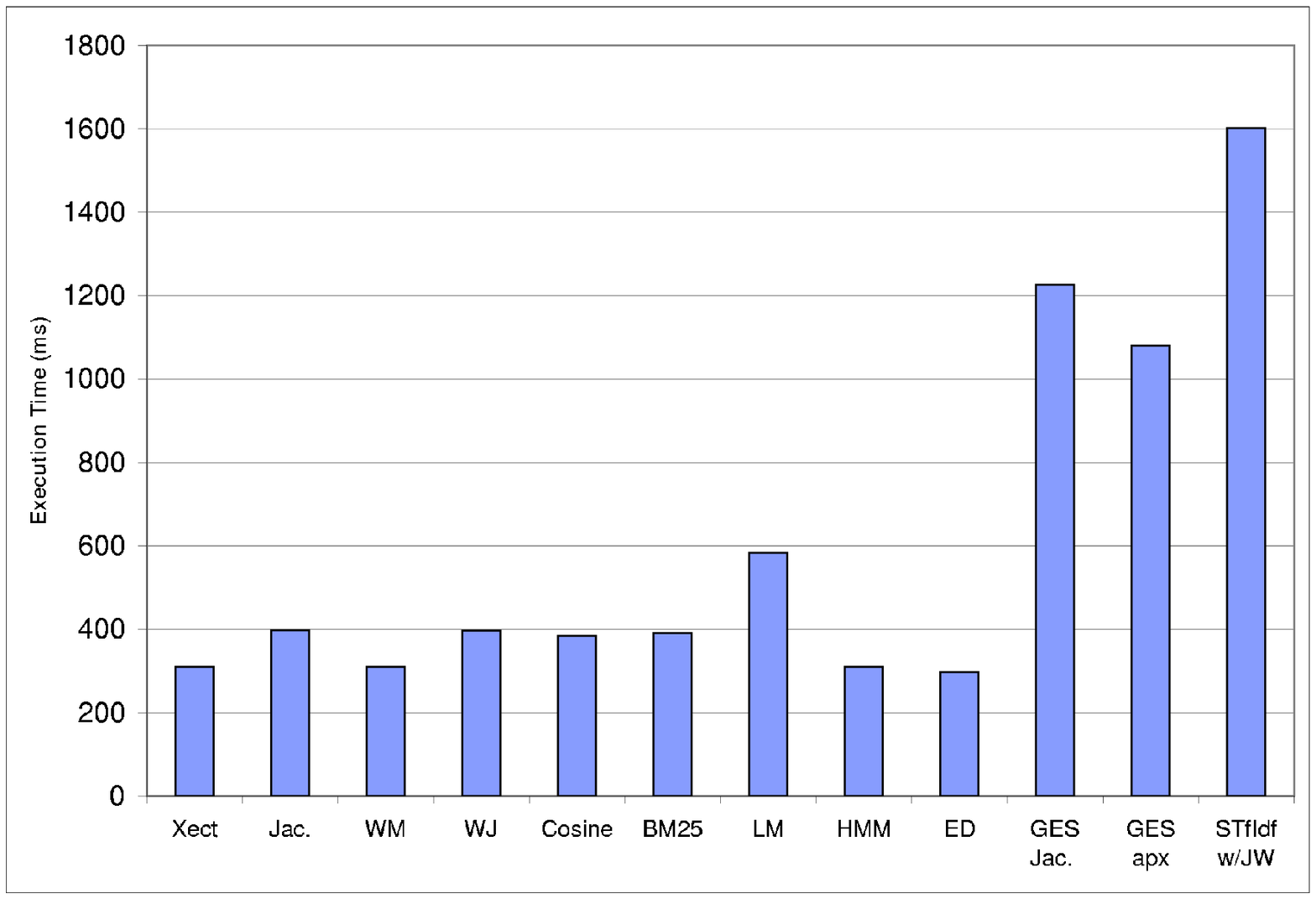,, width=3.5in}
\caption{Query time of different predicates}
\label{querychart}

\end{figure}

Figure \ref{querychart} shows the average query execution time of
different predicates over 100 queries on a table of 10,000 strings with an
average length of 37 characters. The experimental results are consistent with our
analysis. A comparison of the average query time of the predicates shows that
IntersectSize, Jaccard, WeightedMatch, WeightedJaccard,
HMM, BM25 should be among the best since first, they just involve one join and
second, the query token weights do not depend on $idf$ and are easy to compute.
We expect the Cosine predicate to follow these predicates as it has the
additional
overhead of calculating query weights which depend on $idf$ of tokens. The Language
Modeling predicate involves join of 3 tables, so it is comparatively slow.
The GES based predicates are slowest of all since they involve identification of
the best
matching token among the tuples for each query token. $GES^{apx}$ has been
designed to efficiently approximate GES$^{Jaccard}$, so it is expected to be
the fastest of all GES based predicates.  Note that the filtering step of
GES$^{Jaccard}$, GES$^{apx}$ and edit distance require a suitable threshold
$\theta$. Lower value of $\theta$ results
in poor filtering and high post-processing time, while higher value of $\theta$
leads to loss of similar results and hence a drop in accuracy. We used
$\theta$=0.8 for the filtering step in GES$^{Jaccard}$ and GES$^{apx}$ and
$\theta$=0.7 for edit distance, since these
values balance the trade-off between the performance and precision for these
predicates. 
For $GES^{apx}$, we use 5 hash functions for min-hash calculation
(min hash signature of 5).

\subsection{Scalability}
  
\begin{figure}

\centering
\epsfig{file=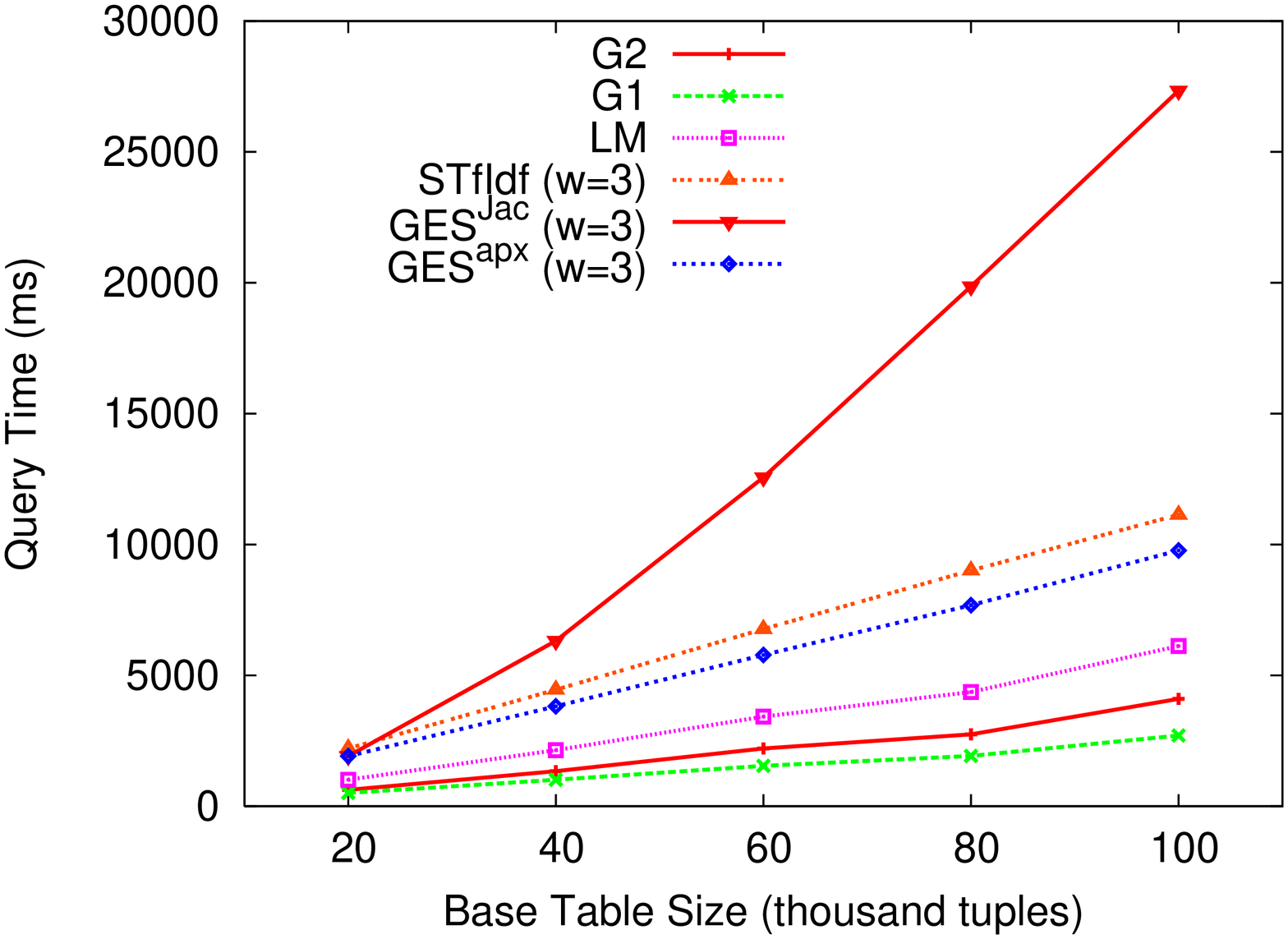,width=3.5in, angle=270}
\caption{\small{Query Time: Variation in Base Table Size}}
\label{scalchart}

\end{figure}

In order to investigate the scalability of our approach, we run experiments on
DBLP datasets with sizes varying from 10k to 100k records. The variation in query time
as the base table size increases is shown in Figure \ref{scalchart}.
The predicates with nearly equal query execution times have been grouped
together.
Group \texttt{G1} includes predicates IntersectSize,
WeightedMatch and
HMM, and the group \texttt{G2} includes  Jaccard, WeightedJaccard, Cosine and
BM25. For predicates other than combination predicates, the results are 
consistent with our analysis of query execution time
presented in Section \ref{qanalysis}. The predicates in group G1 can be thought
of having a weight of 1 for query tokens and they just require a single join to
compute similar tuples. The predicates in group G2 take slightly more time than
predicates in G1 since they have to calculate weights for query tokens. LM
requires join of three tables to get results so it is considerably slower than
predicates in G1 and G2. For the case of combination predicates, query time
depends highly on the value of threshold $\theta$ used for these predicates and 
the number of words in the string. We use the same thresholds we used in Section
\ref{qanalysis} for these predicates. We also limit the size of the
query strings to three words in order to be able to compare the values
among different datasets with other predicates.
 The results show that combination predicates are significantly slower than other predicates since for
each query token, we need to determine the best matching token from the base
tuple using an auxiliary similarity function such as Jaccard and Jaro-Winkler,
apart from the time needed to calculate related weights for word tokens.
GES$^{apx}$ is the fastest in this cluster of predicates.
Increasing the number of words in query strings considerably slows down these
predicates. We excluded edit distance from this experiment because of its
significantly poor accuracy.

\section{Performance Enhancements}
\label{perfEnhance}

 Apart from obvious ways of boosting the performance of algorithms such as
modifying score formulas as described in Section \ref{declarative_chapter}
and building indices on relations to improve execution plans of score
calculation formulas, it is possible to enhance performance of the algorithms
by using filtering and pruning  methods. Filtering based methods try to find a
set of tuples which are promising duplicates by dropping a considerable
percentage of dissimilar records without calculating exact scores.
GES$^{Jaccard}$ is an example of such techniques. Filtering based 
enhancement techniques for declarative framework are described in detail in
\cite{flexiblematchVLDB04}.

\ignore{
Apart from obvious ways of boosting performance of the algorithms such as
tweaking score formulas as described in section \ref{declarative_section}
(declarative framework section of the paper) and building indices on relations
to improve execution plans of score calculation formulas, it is possible to
enhance performance of the algorithms in several other ways. Some of these ways
are based on the fact that to find approximate matches in a database
of several million records, it is not necessary to calculate similarity scores
between the query string and all records since only a few of the records are
potentially duplicates and there are simpler ways of dropping a considerable
percentage of dissimilar records before calculating scores. These types of
enhancements are described in detail in \cite{flexiblematchVLDB04}. 
}

Pruning methods enhance the performance of algorithms mainly based on the nature
of q-grams made out of strings. As results of our experiments suggest, in every
data set, there is a huge number of qgrams that play very little or no role in
the accuracy of predicates. Therefore, a very reasonable policy is to drop those
q-grams in favor of space and running time. One way to do so is to prune weight
tables by dropping tokens with weights less than a threshold during the
preprocessing or the query execution phase. This method can be effective in
some predicates, however, there are two problems associated with this approach.
The first problem is to find 
 the best threshold which depends on the data set and
the predicate. The other is related to the characteristics of some predicates,
specially language modeling and HMM predicates, where dropping tokens will ruin
the
probability distributions calculated for tokens and as a result, the score
formulas would fail to calculate correct similarity scores. 

A better strategy that is shown to be very effective in our experiments is to
prune base relation's tokens table based on IDF of tokens. It is analogous
to the idea of removing \texttt{stopwords} e.g. \texttt{the, an} etc. from the
documents for efficient keyword search. This approach has
several advantages. Although some extra steps will be added to the preprocessing
phase, as a results of pruning, other steps of preprocessing will be
considerably faster and overall, we gain substantial performance
improvement in the preprocessing phase (except for unweighted naive predicates
where there is no step other than preprocessing). Since all weights are
calculated from the pruned tokens table, the probability distributions of tokens
will remain meaningful. The benefit for query execution time and the effect on
accuracy depends on the threshold used for pruning.
 
\begin{figure*}
\centering
\begin{tabular}{|cc|} \hline
\epsfig{file=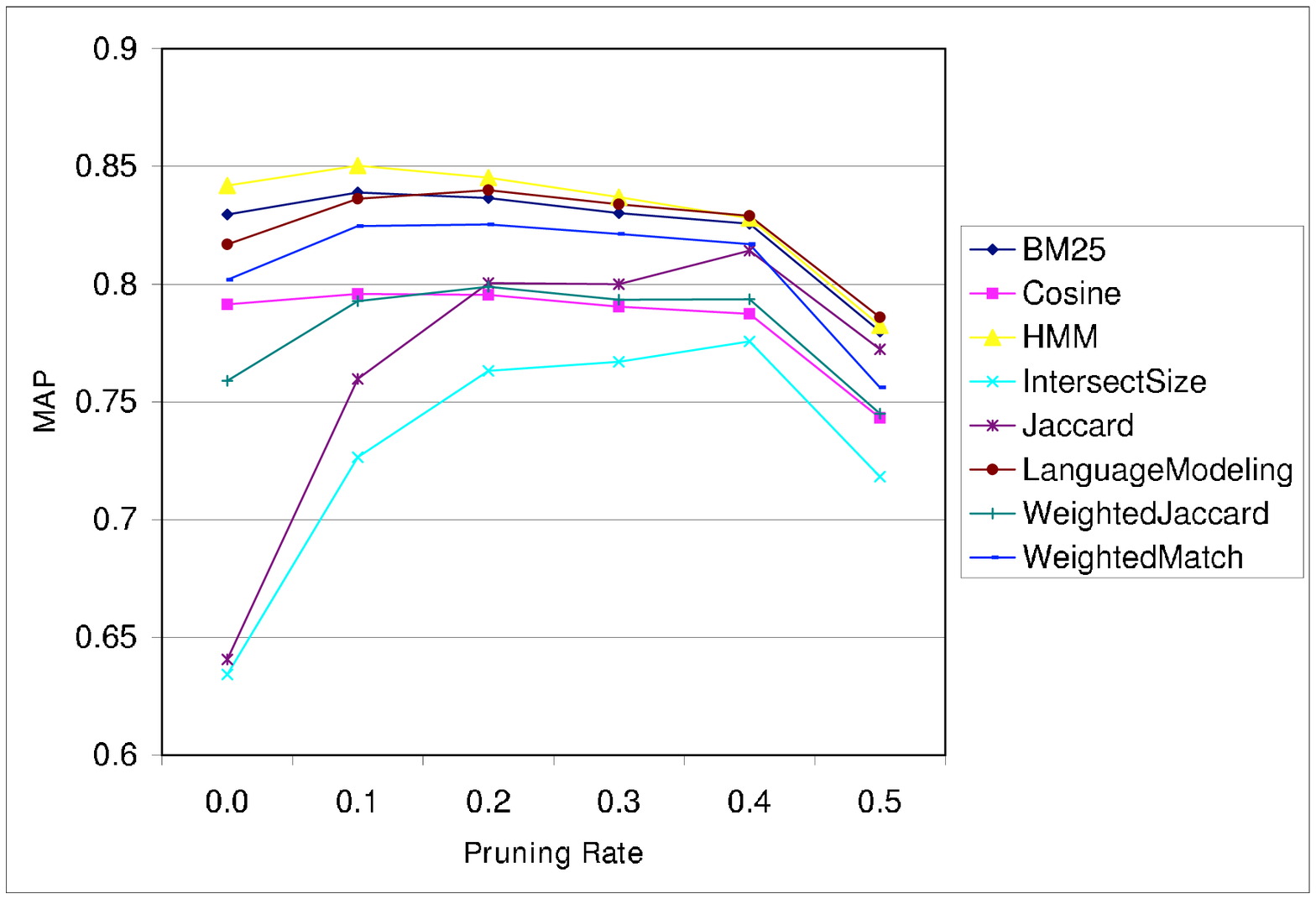,width=3.2in}
 &
\epsfig{file=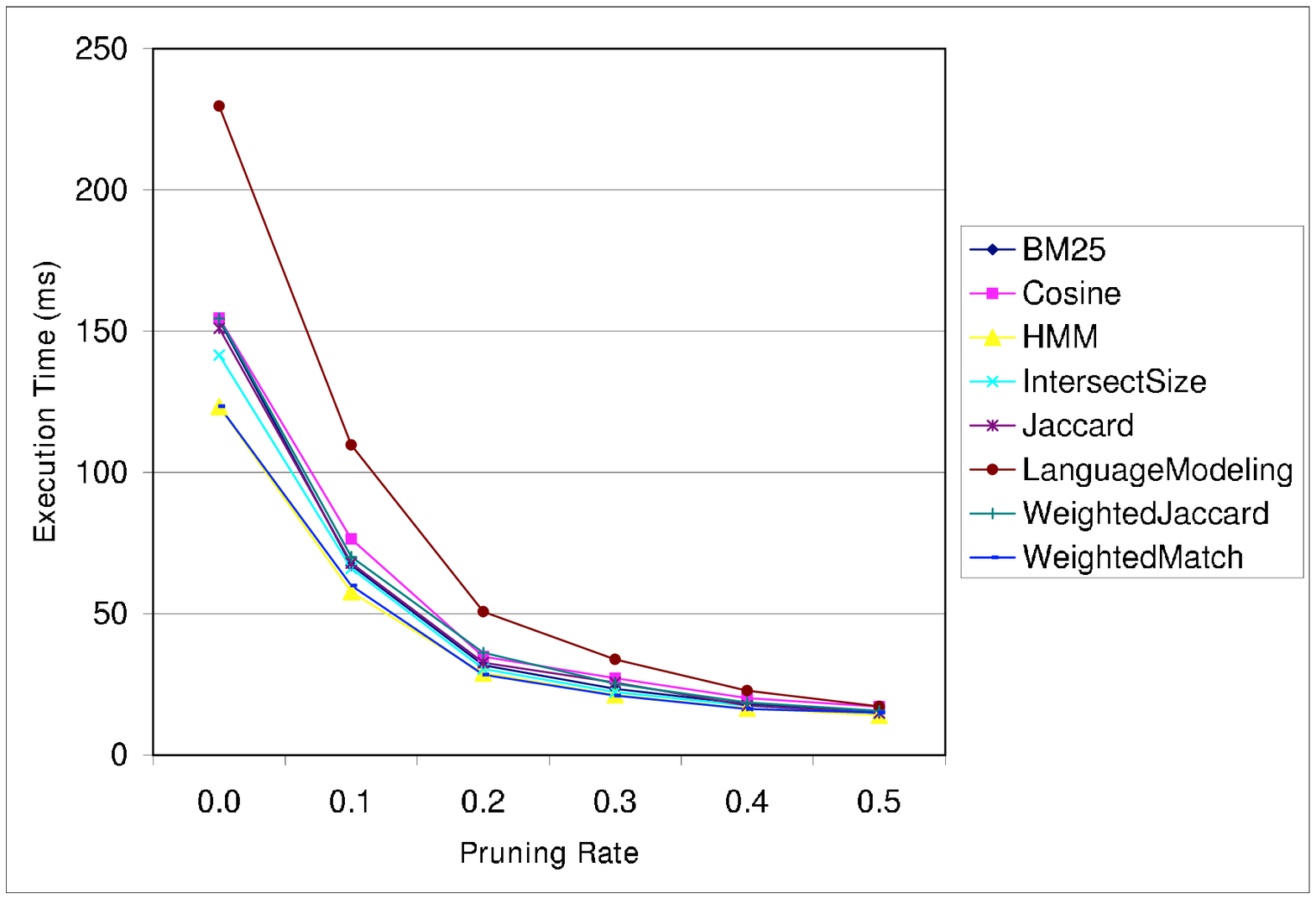,width=3.2in} \\
 (a) MAP vs. Pruning Rate & (b) Execution Time vs. Pruning Rate \\ 
\hline
\end{tabular}
\caption{Effect of Pruning on MAP and Execution Time of Different predicates} 
\label{prunechart1}
\end{figure*}

Figure \ref{prunechart1} shows the effect of threshold used for pruning on MAP
and execution time for a dirty data set of company names. We use a threshold in
the form of $MIN(idf) + rate * (MAX(idf)-MIN(idf))$ and change the rate from $0$
(i.e., no pruning) to $0.5$. As we can see in the figure, in the unweighted
naive predicates, pruning results in a considerable gain in accuracy. This is
obviously due to dropping tokens with low IDF from strings, so the similarity
score for these predicates will be intersection and Jaccard coefficient of only
important tokens. Interestingly, with low threshold values, accuracy of all
other predicates benefit from pruning. This shows that tokens with very low IDF
do
not play an effective role in the similarity scores. For our example data set, a
pruning rate between $0.2$ and $0.3$ has the best effect on both accuracy and a
remarkable effect on performance, while rate up to $0.4$ results in reasonable
drop in accuracy and little more gain in performance.
 
\ignore{ 
Note that in IR literature, pruning is used just to improve performance, but
we also get accuracy boost by setting suitable threshold for pruning. In IR,
only stopwords can not make two documents look similar, but for our case the
low IDF tokens in fact contribute negatively to accuracy by making two different
tuples look similar. Consider the query \texttt{IBM inc} which aim to find the
company names containing \texttt{IBM} in it, but most of the predicates give
same
rank to the erroneous tuple \texttt{BIM inp} and \texttt{ATT inc}. Dropping the
stopword \texttt{inc} will remove this ambiguity and put \texttt{BIM inp} ahead
of \texttt{ATT}. This example also shows that it is important to set the
suitable threshold just to remove all stopwords without removing any of the
important tokens.
}

\begin{figure}
\centering
\epsfig{file=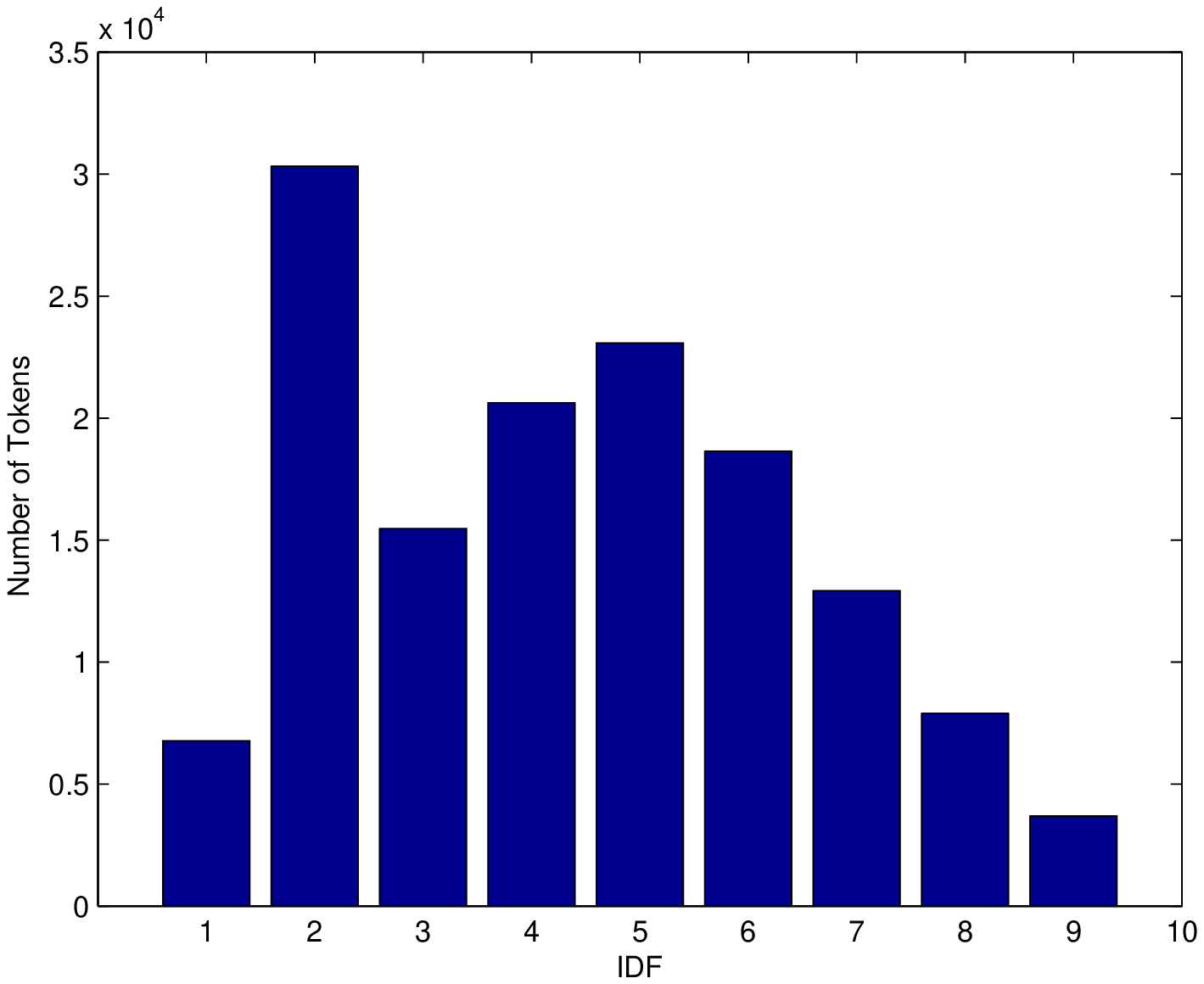,width=3.2in}
 \caption{IDF distribution of qgrams of size 3 for CU1 data set} 
\label{idf-chart}
\end{figure}

To further investigate the effect of pruning based on IDF on performance and
accuracy of predicates, we examine the IDF distribution in  our data sets.
Figure
\ref{idf-chart} shows the distribution of IDF weights for CU1 data set. The
distribution is similar in all other data sets. As it can be seen, there is a huge
number of tokens with low IDF. For this data set, a pruning rate of $0.33$ will
drop nearly 150,000 out of 250,000 tokens which results in a huge performance
gain and a
very little drop in accuracy of predicates (except unweighted naive predicates
that
benefit from pruning as described above).

\ignore{
of company
names with high and low error rates. The lines in the figure show the threshold
rate $0.33$. As it can be seen, this threshold value results in dropping nearly
150,000 out of 250,000 tokens which results in a huge performance gain and a
very little drop in accuracy of predicates (except unweighted naive predicates
that
benefit from pruning as described above).
}
 
\ignore{ 
 In a number of predicates, it is possible to prune weights table in order to
save
space and execution time. Pruning can be performed during the preprocessing or
the query execution phase by simply dropping from weights table, those tokens
with weights less than a specific threshold. Increasing threshold will results
in lower accuracy but faster query execution time.

  Figure \ref{prunechart1} shows the effect of pruning on accuracy and
performance of cosine and BM25 predicates on different data sets. The pruning
threshold is chosen to be average of the weight values of tokens minus half of
their standard deviation. As it can be seen in the figure, this threshold
results in 3\% loss of accuracy in dirty data sets and less than 1\% loss of
accuracy in medium and lower error data sets for both predicates, while the
execution time of the pruned versions are up to 7 times faster then the original
predicates.
}

\ignore{ SHORT VERSION:
Apart from ways of boosting the performance of algorithms such as
modifying score formulas as described in Section \ref{declarative_section}
and building indices on relations to improve execution plans of score
calculation formulas, it is possible to enhance performance of the algorithms
by using filtering and pruning  methods. Filtering based methods try to find a
set of tuples which are promising duplicates by dropping a considerable
percentage of dissimilar records without calculating exact scores.
GES$^{Jaccard}$ is an example of such techniques. Filtering based 
enhancement techniques for declarative framework are described in detail in
\cite{flexiblematchVLDB04}.

Pruning based methods drop a significant percentage of tokens which play very
little or no role in deciding the similarity of tuples. It is analogous to the
idea of removing \texttt{stopwords} e.g. \texttt{the, an} etc. from the
documents for efficient keyword search. We prune the tokens of a base relation
based on $idf$ of tokens. We drop all the tokens which have $idf$ less than
threshold $\theta$. Note that $\theta$ has to be chosen carefully since low
value of $\theta$ does not lead to any performance gain where as high value of
$\theta$ may drop important tokens and lead to drop in accuracy. Pruning tokens
based on $idf$ improves performance of both the preprocessing step and the query
execution step. Our experiments show that  we can reduce the query execution
time of the qgram based predicates by a factor of five. In addition to
performance gain, low threshold value also improves accuracy by removing
irrelevant tokens. The detailed results and discussion can be found in the
extended version of this paper \cite{extendedVersion}.
}

\section{Summary of Evaluation}

We presented an exhaustive evaluation of approximate selection predicates by 
grouping them into five classes based on their characteristics:
overlap predicates, aggregate weighted predicates, edit-based predicates,
combination predicates and language modeling predicates. We experimentally show
how predicates in each of these classes perform in terms of accuracy,
preprocessing and execution time. Within our framework, the overlap predicates
are relatively efficient but have low accuracy. Edit based predicates perform
worse in terms of accuracy but are relatively fast due to the filtering step they
employ. The aggregate weighted predicates, specifically BM25, perform very well both in
terms of accuracy and efficiency. Both the predicates from the language
modeling cluster perform well in terms of accuracy. Moreover, HMM is as fast as
simple overlap predicates. The combination predicates are considerably slow due
to their two levels of tokenization. Among the combination predicates, GES based
predicates are robust in handling edit errors but fail considerably in
capturing token swap errors. SoftTFIDF with Jaro-Winkler performs nearly equal
to BM25 and HMM and is among the best in terms of accuracy, although it is
the slowest predicate. This establishes the effectiveness of BM25 and HMM
predicates for approximate matching in large databases. 

\chapter{Conclusions}
We proposed new similarity predicates for approximate selections 
based on probabilistic information retrieval and
presented their declarative instantiation. We presented an in-depth comparison of
accuracy and performance of these new predicates along with
existing predicates, grouping them into classes based on their primary
characteristics. Our experiments show that the new predicates are both
effective as well as efficient for data cleaning applications.



%
%

\addcontentsline{toc}{chapter}{Bibliography}
\bibliographystyle{plain}

\begin{thebibliography}{10}

\bibitem{ananthakrishna02eliminating}
R.~Ananthakrishna, S.~Chaudhuri, and V.~Ganti.
\newblock Eliminating fuzzy duplicates in data warehouses.
\newblock In {\em Proceedings of the 28th International Conference on Very
  Large Databases (VLDB)}, 2002.

\bibitem{SSJOIN2}
Arvind Arasu, Venkatesh Ganti, and Raghav Kaushik.
\newblock Efficient exact set-similarity joins.
\newblock In {\em Proceedings of the 32nd international conference on
  Very large data bases (VLDB)}, pages 918--929. VLDB Endowment, 2006.

\bibitem{broder00minwise}
Andrei~Z. Broder, Moses Charikar, Alan~M. Frieze, and Michael Mitzenmacher.
\newblock Min-wise independent permutations.
\newblock {\em Journal of Computer and System Sciences}, 60(3):630--659, 2000.

\bibitem{fmsSIGMOD03}
Surajit Chaudhuri, Kris Ganjam, Venkatesh Ganti, and Rajeev Motwani.
\newblock Robust and efficient fuzzy match for online data cleaning.
\newblock In {\em SIGMOD'03: Proceedings of the 2003 ACM SIGMOD International Conference
  on Management of Data}, pages 313--324, June 2003.

\bibitem{SSJOIN}
Surajit Chaudhuri, Venkatesh Ganti, and Raghav Kaushik.
\newblock A primitive operator for similarity joins in data cleaning.
\newblock In {\em ICDE '06: Proceedings of the 22nd International Conference on
  Data Engineering (ICDE)}, page~5, Washington, DC, USA, 2006. IEEE Computer
  Society.

\bibitem{Cohen98}
William~W. Cohen.
\newblock Integration of heterogeneous databases without common domains using
  queries based on textual similarity.
\newblock In {\em SIGMOD '98: Proceedings of the 1998 ACM SIGMOD international
  conference on Management of data}, pages 201--212, New York, NY, USA, 1998.
  ACM Press.

\bibitem{cohen1}
William~W. Cohen, Pradeep Ravikumar, and Stephen~E. Fienberg.
\newblock A comparison of string distance metrics for name-matching tasks.
\newblock In {\em Proceedings of IJCAI-03 Workshop on Information Integration
  on the Web (IIWeb-03)}, pages 73--78, 2003.

\bibitem{linkage90}
J.~B. Copas and F.~J. Hilton.
\newblock Record linkage: statistical models for matching computer records.
\newblock {\em Journal of the Royal Statistical Society}, pages 287--320, 1990.

\bibitem{Fellegi69}
Ivan~P. Fellegi and Alan~B. Sunter.
\newblock A theory for record linkage.
\newblock {\em Journal of the American Statistical Association},
  64(328):1183--1210, 1969.

\bibitem{Galhardas01}
Helena Galhardas, Daniela Florescu, Dennis Shasha, Eric Simon, and
  Cristian-Augustin Saita.
\newblock Declarative data cleaning: Language, model, and algorithms.
\newblock In {\em Proceedings of the International Conference on Very
  Large Databases (VLDB)}, pages 371--380, 2001.

\bibitem{joinForFreeVLDB01}
Luis Gravano, Panagiotis~G. Ipeirotis, H.~V. Jagadish, Nick Koudas,
  S.~Muthukrishnan, and Divesh Srivastava.
\newblock Approximate string joins in a database (almost) for free.
\newblock In {\em Proceedings of the 27th International Conference on
  Very Large Data Bases (VLDB)}, pages 491--500, San Francisco, CA, USA, 2001. Morgan
  Kaufmann Publishers Inc.

\bibitem{textjoinWWW03}
Luis Gravano, Panagiotis~G. Ipeirotis, Nick Koudas, and Divesh Srivastava.
\newblock Text joins for data cleansing and integration in an rdbms.
\newblock In {\em Proceedings of the 19th International Conference on Data
  Engineering (ICDE)}, pages 729--731, March 2003.

\bibitem{GusfieldBook}
Dan Gusfield.
\newblock {\em Algorithms on strings, trees, and sequences: computer science
  and computational biology}.
\newblock Cambridge University Press, New York, NY, USA, 1997.

\bibitem{MergePurge}
Mauricio~A. Hern\'andez and Salvatore~J. Stolfo.
\newblock Real-world data is dirty: Data cleansing and the merge/purge problem.
\newblock {\em Data Min. Knowl. Discov.}, 2(1):9--37, 1998.

\bibitem{Jaro84}
M.~A. Jaro.
\newblock Advances in record linkage methodology as applied to matching the
  1985 census of tampa.
\newblock {\em Journal of the American Statistical Association}, pages
  414--420, 1984.

\bibitem{flexiblematchVLDB04}
Nick Koudas, Amit Marathe, and Divesh Srivastava.
\newblock Flexible string matching against large databases in practice.
\newblock In {\em Proceedings of the International Conference on Very
  Large Databases (VLDB)}, pages 1078--1086, August 2004.

\bibitem{SpiderSIGMOD05}
Nick Koudas, Amit Marathe, and Divesh Srivastava.
\newblock Spider: flexible matching in databases.
\newblock In {\em SIGMOD '05: Proceedings of the 2005 ACM SIGMOD international
  conference on Management of data}, pages 876--878, New York, NY, USA, 2005.
  ACM Press.

\bibitem{Koudas06Spider}
Nick Koudas, Amit Marathe, and Divesh Srivastava.
\newblock Using spider: an experience report.
\newblock In {\em SIGMOD '06: Proceedings of the 2006 ACM SIGMOD international
  conference on Management of data}, page 719, 2006.

\bibitem{Koudas06}
Nick Koudas, Sunita Sarawagi, and Divesh Srivastava.
\newblock Record linkage: similarity measures and algorithms.
\newblock In {\em SIGMOD '06: Proceedings of the 2006 ACM SIGMOD international
  conference on Management of data}, pages 802--803, 2006.

\bibitem{Koudas05}
Nick Koudas and Divesh Srivastava.
\newblock Approximate joins: Concepts and techniques.
\newblock In {\em Proceedings of the International Conference on Very
  Large Databases (VLDB)}, page 1363, 2005.

\bibitem{hmmSIGIR99}
David R.~H. Miller, Tim Leek, and Richard~M. Schwartz.
\newblock A hidden markov model information retrieval system.
\newblock In {\em Proceedings of the 22nd Annual International ACM SIGIR
  Conference on Research and Development in Information Retrieval}, pages
  214--221, August 1999.

\bibitem{ponte98}
Jay~M. Ponte and W.~Bruce Croft.
\newblock A language modeling approach to information retrieval.
\newblock In {\em Proceedings of the 21st Annual International ACM SIGIR
  Conference on Research and Development in Information Retrieval}, pages
  275--281, August 1998.

\bibitem{rabiner89}
L.R. Rabiner.
\newblock A tutorial on hidden markov models and selected applications inspeech
  recognition.
\newblock In {\em Proceedings of the IEEE}, volume~77, pages 257--286, 1989.

\bibitem{robertson04}
Stephen Robertson.
\newblock Understanding inverse document frequency: on theoretical arguments.
\newblock {\em Journal of Documentation}, 60(5):503--520, 2004.

\bibitem{bm25TREC95}
Stephen~E. Robertson, Steve Walker, Micheline Hancock-Beaulieu, Mike Gatford,
  and A.~Payne.
\newblock Okapi at trec-4.
\newblock In {\em TREC}, 1995.

\bibitem{Salton88}
Gerard Salton and Chris Buckley.
\newblock Term-weighting approaches in automatic text retrieval.
\newblock {\em Information Processing and Management}, 24(5):513--523, 1988.

\bibitem{irbook}
Gerard Salton and Michael~J. McGill.
\newblock {\em Introduction to Modern Information Retrieval}.
\newblock McGraw-Hill, Inc., New York, NY, USA, 1986.

\bibitem{sunitaSIGMOD04}
Sunita Sarawagi and Alok Kirpal.
\newblock Efficient set joins on similarity predicates.
\newblock In {\em SIGMOD'04: Proceedings of the ACM SIGMOD International Conference on
  Management of Data}, pages 743--754, June 2004.

\bibitem{Winkler99}
W.~E. Winkler.
\newblock The state of record linkage and current research problems.
\newblock Technical Report RR99/04, US Bureau of the Census, 1999.

\end{thebibliography}


\newpage

\appendix


\chapter{Data preparation SQL Statements}
\label{ap_1}

 We assume the base relation \texttt{BASE\_TABLE} has an integer tuple id attribute \texttt{tid} and a string valued attribute \texttt{string}. The following SQL statements tokenize the base relation, creating \texttt{BASE\_TOKENS(tid, token)}. Assuming that the query relation \texttt{QUERY\_TABLE} has a single string valued attribute \texttt{string}, the same SQL statements can be used for tokenization of the query string by removing \texttt{tid} from the statements.

\section{Qgram generation}


{
\scriptsize \tt  
\begin{tabular}{|l|} \hline
-- MAX\_STR\_SIZE is the maximum string length and $q$ is the size of the qgrams.\\
INSERT INTO INTEGERS(i) VALUES (1), (2), \dots, (MAX\_STR\_SIZE + $(q-1)$) \\
\\
INSERT INTO BASE\_TOKENS(tid, token) \\
SELECT \hspace{0.01in}     tid, SUBSTRING(CONCAT(SUBSTRING(`\$\dots\$`,1,$q$-1), \\
       \hspace{0.40in}     UPPER( REPLACE(CONCAT(string),` `,SUBSTRING(`\$\dots\$`,1,$q$-1))), \\
       \hspace{0.40in}     SUBSTRING(`\$\dots\$`,1,$q$-1)) , INTEGERS.i, $q$)    \\
FROM   \hspace{0.11in}     INTEGERS INNER JOIN BASE\_TABLE ON \\
       \hspace{0.40in}     INTEGERS.i <= LENGTH( REPLACE(CONCAT(string),` `,SUBSTRING(`\$\dots\$`,1,$q$-1))) + ($q$-1)
 \\ \hline
\end{tabular}
}


\section{Word token generation}

{
\scriptsize \tt  
\begin{tabular}{|ll|} \hline
\multicolumn{2}{|l|}{INSERT INTO BASE\_TOKENS(tid, token) } \\
SELECT & tid, SUBSTRING(CONCAT(string), 1, LOCATE(' ', CONCAT(string)) - 1) \\
FROM   & BASE\_TABLE \\
WHERE  & LOCATE(' ', CONCAT(string)) > 0 \\
\multicolumn{2}{|l|}{UNION ALL} \\
SELECT & tid, SUBSTRING(CONCAT(string), N1.I+1, N2.I - N1.I-1) \\
FROM   & BASE\_TABLE, INTEGERS N1, INTEGERS N2 \\
WHERE  & N1.I = LOCATE(' ', CONCAT(string), N1.I) AND N2.I = LOCATE(' ', CONCAT(string), N1.I + 1) \\
\multicolumn{2}{|l|}{UNION ALL} \\
SELECT & tid, SUBSTRING(CONCAT(string), LENGTH(CONCAT(string)) - LOCATE(' ', REVERSE(CONCAT(string)))+2) \\
FROM   & BASE\_TABLE \\
WHERE  & LOCATE(' ', CONCAT(string)) > 0 \\
\multicolumn{2}{|l|}{UNION ALL} \\
SELECT & tid, CONCAT(string) \\
FROM   & BASE\_TABLE \\
WHERE  & LOCATE(' ', CONCAT(string)) = 0 \\ \hline
\end{tabular}
}

\section{Qgram generation of the word tokens (for combination predicates)}

{
\scriptsize \tt  
\begin{tabular}{|ll|} \hline
\multicolumn{2}{|l|}{INSERT INTO BASE\_QGRAMS(tid, token, qgram)} \\
SELECT & tid, token, \\
       & SUBSTRING(CONCAT(SUBSTRING(`\$\dots\$`,1,$q$-1), UPPER(token), SUBSTRING(`\$\dots\$`,1,$q$-1)), INTEGERS.I, $q$) \\
FROM   & INTEGERS INNER JOIN BASE\_TOKENS ON INTEGERS.I <= LENGTH(token) + ($q$-1)  \\
GROUP BY & tid, token, qgram \\ \hline
\end{tabular}
}

\chapter{SQL Statements for Predicates}
\label{ap_2}

\section{Overlap Predicates}

\subsection{IntersectSize}
{
\scriptsize \tt  
\begin{tabular}{|ll|} \hline
\multicolumn{2}{|c|}{Query} \\ \hline
\multicolumn{2}{|l|}{INSERT INTO INTERSECT\_RESULTS(tid, score)} \\
SELECT   & R1.tid, COUNT(*)\\
FROM     & BASE\_TOKENS R1, QUERY\_TOKENS R2 \\
WHERE    & R1.token = R2.token \\
GROUP BY & R1.tid \\
\hline
\end{tabular}
}

\subsection{Jaccard}
{
\scriptsize \tt  
\begin{tabular}{|ll|} \hline
\multicolumn{2}{|c|}{Preprocessing} \\ \hline
\multicolumn{2}{|l|}{ INSERT INTO BASE\_DDL(tid, ddl) } \\
SELECT & T.tid, COUNT(*) \\
FROM   & BASE\_TOKENS T \\
GROUP BY & T.tid \\
& \\
\multicolumn{2}{|l|}{INSERT INTO BASE\_TOKENSDDL(tid, token, ddl) } \\
SELECT & T.tid, T.token, D.ddl \\
FROM   & BASE\_TOKENS T, BASE\_DDL D \\
WHERE  & T.tid = D.tid \\ \hline
\multicolumn{2}{|c|}{Query} \\ \hline
\multicolumn{2}{|l|}{INSERT INTO JACCARD\_RESULTS(tid, score) } \\
SELECT   & S1.tid, COUNT(*)/(S1.ddl + S2.ddl - COUNT(*)) \\
FROM     & BASE\_TOKENSDDL S1, QUERY\_TOKENS R2, (SELECT COUNT(*) AS ddl FROM QUERY\_TOKENS T) S2 \\
WHERE    & S1.token = R2.token \\
GROUP BY & S1.tid \\
\hline
\end{tabular}
}

\newpage

\subsection{WeightedMatch}

{
\scriptsize \tt  
\begin{tabular}{|ll|} \hline
\multicolumn{2}{|c|}{Preprocessing} \\ \hline
\multicolumn{2}{|l|}{INSERT INTO BASE\_SIZE(size)} \\
SELECT & COUNT(*) \\
FROM   & BASE\_TABLE \\
& \\
\multicolumn{2}{|l|}{INSERT INTO BASE\_TF(tid, token, tf)} \\
SELECT & T.tid, T.token, COUNT(*) \\
FROM   & BASE\_TOKENS T \\
GROUP BY & T.tid, T.token\\
& \\
\multicolumn{2}{|l|}{INSERT INTO BASE\_BMIDF(token, midf)} \\
SELECT   & T.token, LOG(S.SIZE - COUNT(T.tid) + 0.5) - LOG(COUNT(T.tid) + 0.5) \\
FROM     & BASE\_TF T, BASE\_SIZE S \\
GROUP BY & T.token\\
& \\
\multicolumn{2}{|l|}{INSERT INTO BASE\_WEIGHTS(tid, token, weight)} \\
SELECT & T.tid, T.token, I.midf \\
FROM   & BASE\_BMIDF I, BASE\_TF T \\
WHERE  & I.token = T.token \\
\hline
\multicolumn{2}{|c|}{Query} \\ \hline
\multicolumn{2}{|l|}{INSERT INTO WEIGHTEDMATCH\_RESULTS(tid1, tid2, score)} \\
SELECT   & W1.tid, T2.tid, SUM(W1.weight) \\
FROM     & BASE\_WEIGHTS W1, QUERY\_TOKENS T2 \\
WHERE    & W1.token = T2.token \\
GROUP BY & T2.tid, W1.tid \\ \hline 
\end{tabular}
}

\subsection{WeightedJaccard}

{
\scriptsize \tt  
\begin{tabular}{|ll|} \hline
\multicolumn{2}{|c|}{Preprocessing} \\ \hline
\multicolumn{2}{|l|}{INSERT INTO BASE\_SIZE(size)} \\
SELECT & COUNT(*) \\
FROM   & BASE\_TABLE\\
& \\
\multicolumn{2}{|l|}{INSERT INTO BASE\_TF(tid, token, tf)} \\
SELECT & T.tid, T.token, COUNT(*) \\
FROM   & BASE\_TOKENS T \\
GROUP BY & T.tid, T.token\\
& \\
\multicolumn{2}{|l|}{INSERT INTO BASE\_BMIDF(token, midf)} \\
SELECT   & T.token, LOG(S.SIZE - COUNT(T.tid) + 0.5) - LOG(COUNT(T.tid) + 0.5) \\
FROM     & BASE\_TF T, BASE\_SIZE S \\
GROUP BY & T.token\\
& \\
\multicolumn{2}{|l|}{INSERT INTO BASE\_WEIGHTS(tid, token, weight)} \\
SELECT & T.tid, T.token, I.midf \\
FROM   & BASE\_BMIDF I, BASE\_TOKENS T \\
WHERE  & I.token = T.token\\
& \\
\multicolumn{2}{|l|}{INSERT INTO BASE\_DDL(tid, ddl)} \\
SELECT & W.tid, SUM(weight) \\
FROM   & BASE\_WEIGHTS W \\
GROUP BY & W.tid\\
& \\
\multicolumn{2}{|l|}{INSERT INTO BASE\_TOKENSDDL(tid, token, ddl, weight)} \\
SELECT & W.tid, W.token, D.DDL, W.WEIGHT \\
FROM   & BASE\_WEIGHTS W, BASE\_DDL D \\
WHERE  & W.tid = D.tid\\
\hline
\multicolumn{2}{|c|}{Query} \\ \hline
\multicolumn{2}{|l|}{INSERT INTO WJ\_RESULTS(tid, score)} \\
SELECT & S1.tid, SUM(S1.weight)/(S1.ddl + S2.ddl - SUM(S1.weight))  \\
FROM  & BASE\_TOKENSDDL S1, QUERY\_TOKENS R2, \\
      &                ( SELECT SUM(T.weight) AS ddl \\
      & \hspace{0.07in} FROM (SELECT T.token, I.IDF AS weight \\
      & \hspace{0.38in}       FROM BASE\_IDF I, QUERY\_TOKENS T \\
      & \hspace{0.38in}       WHERE I.token = T.token) T ) S2 \\
WHERE & S1.token = R2.token \\
GROUP & BY S1.tid \\
\hline
\end{tabular}
}

\newpage

\section{Aggregate Weighted Predicates}

\subsection{Tfidf Cosine Predicate}
{
\scriptsize \tt  
\begin{tabular}{|ll|} \hline
\multicolumn{2}{|c|}{Preprocessing} \\ \hline
\multicolumn{2}{|l|}{INSERT INTO BASE\_SIZE(size)} \\
SELECT & COUNT(*) \\
FROM   & BASE\_TABLE\\
& \\
\multicolumn{2}{|l|}{INSERT INTO BASE\_IDF(token, idf)} \\
SELECT & T.token, LOG(S.SIZE) - LOG(COUNT(DISTINCT T.tid)) \\
FROM   & BASE\_TOKENS T, BASE\_SIZE S \\
GROUP BY & T.token\\
& \\
\multicolumn{2}{|l|}{INSERT INTO BASE\_TF(tid, token, tf)} \\
SELECT & T.tid, T.token, COUNT(*) \\
FROM   & BASE\_TOKENS T \\
GROUP BY & T.tid, T.token\\
& \\
\multicolumn{2}{|l|}{INSERT INTO BASE\_LENGTH(tid, len)} \\
SELECT &  T.tid, SQRT(SUM(I.idf*I.idf*T.tf*T.tf)) \\
FROM   &  BASE\_IDF I, BASE\_TF T \\
WHERE  &  I.token = T.token \\
GROUP BY & T.tid\\
& \\
\multicolumn{2}{|l|}{INSERT INTO BASE\_WEIGHTS(tid, token, weight)} \\
SELECT & T.tid, T.token, I.idf*T.tf/L.len \\
FROM   & BASE\_IDF I, BASE\_TF T, BASE\_LENGTH L \\
WHERE  & I.token = T.token AND T.tid = L.tid\\
\hline
\end{tabular}

\begin{tabular}{|ll|} \hline
\multicolumn{2}{|c|}{Query} \\ \hline
\multicolumn{2}{|l|}{INSERT INTO COSINE\_RESULTS(tid, score)} \\
SELECT   & R1W.tid, SUM(R1W.weight*R2W.weight) \\
FROM     & BASE\_WEIGHTS R1W, \\
         & (SELECT T.token, QIDF.idf*QTF.tf/QLEN.length AS weight \\
         &  FROM (SELECT R.token, R.idf \\
         &  \hspace{0.33in}      FROM QUERY\_TOKENS S, BASE\_IDF R \\
         &  \hspace{0.33in}      WHERE S.token = R.token \\
         &  \hspace{0.33in}      GROUP BY S.token) QIDF, \\
         &  \hspace{0.30in}     (SELECT T.token, COUNT(*) AS tf \\
         &  \hspace{0.33in}      FROM QUERY\_TOKENS T \\
         &  \hspace{0.33in}      GROUP BY T.tid, T.token) QTF, \\
         &  \hspace{0.30in}     (SELECT SQRT(SUM(QIDF.idf*QIDF.idf*QTF.tf*QTF.tf)) AS length \\
         &  \hspace{0.33in}      FROM (SELECT R.token, R.idf \\
         &  \hspace{0.73in}            FROM QUERY\_TOKENS S, BASE\_IDF R \\
         &  \hspace{0.73in}            WHERE S.token = R.token \\
         &  \hspace{0.73in}            GROUP BY S.token) QIDF, \\
         &  \hspace{0.70in}           (SELECT T.token, COUNT(*) AS tf \\
         &  \hspace{0.73in}            FROM QUERY\_TOKENS T \\
         &  \hspace{0.73in}            GROUP BY T.token) QTF \\
         &  \hspace{0.33in}      WHERE I.token = T.token) QLEN \\
         &  WHERE QIDF.token = QTF.token) R2W \\
WHERE    & R1W.token = R2W.token \\
GROUP BY & R1W.tid \\
\hline
\end{tabular}
}

\newpage

\subsection{BM25 Predicate}

{
\scriptsize \tt  
\begin{tabular}{|ll|} \hline
\multicolumn{2}{|c|}{Preprocessing} \\ \hline
\multicolumn{2}{|l|}{INSERT INTO BASE\_SIZE(size)} \\
SELECT & COUNT(*) \\
FROM   & BASE\_TABLE\\
& \\
\multicolumn{2}{|l|}{INSERT INTO BASE\_TF(tid, token, tf)} \\
SELECT   & T.tid, T.token, COUNT(*) \\
FROM     & BASE\_TOKENS T \\
GROUP BY & T.tid, T.token\\
& \\
\multicolumn{2}{|l|}{INSERT INTO BASE\_BMIDF(token, midf)} \\
SELECT   & T.token, LOG(S.SIZE - COUNT(T.tid) + 0.5) - LOG(COUNT(T.tid) + 0.5) \\
FROM     & BASE\_TF T, BASE\_SIZE S \\
GROUP BY & T.token\\
& \\
\multicolumn{2}{|l|}{INSERT INTO BASE\_BMBASELENGTH(tid, len)} \\
SELECT   & T.tid, SUM(T.tf) \\
FROM     & BASE\_TF T \\
GROUP BY & T.tid\\
& \\
\multicolumn{2}{|l|}{INSERT INTO BASE\_BMBASEAVGLENGTH(avglen)} \\
SELECT & AVG(len) \\
FROM   & BASE\_BMBASELENGTH\\
& \\
\multicolumn{2}{|l|}{INSERT INTO BASE\_BMBASEMODTF(tid, token, mtf)} \\
SELECT & T.tid, T.token, (T.tf*($k_1$+1)) / ( ((( 1 - $b$)+($b$*L.DL/A.AVGDL))*$k_1$) + T.tf ) \\
FROM   & BASE\_BMBASELENGTH L, BASE\_BMBASEAVGLENGTH A,BASE\_TF T \\
WHERE  & L.tid = T.tid\\
& \\
\multicolumn{2}{|l|}{INSERT INTO BASE\_BMBASEWEIGHTS(tid, token, weight)} \\
SELECT & T.tid, T.token, T.mtf*I.midf \\
FROM   & BASE\_BMBASEMODTF T, BASE\_BMIDF I \\
WHERE  & T.token = I.token \\ \hline
\multicolumn{2}{|c|}{Query} \\ \hline
\multicolumn{2}{|l|}{INSERT INTO BM25\_RESULTS(tid, score)}\\
SELECT   & B.tid, SUM(B.weight * S.mtf) \\
FROM     & BASE\_BMBASEWEIGHTS B, \\
         & \hspace{0.00in} (SELECT token, (COUNT(*))*($k_3$+1) / ($k_3$+COUNT(*)) AS mtf \\
         & \hspace{0.03in}  FROM QUERY\_TOKENS T \\
         & \hspace{0.03in}  GROUP BY T.token) S \\
WHERE    & B.token = S.token \\
GROUP BY & B.tid \\
\hline
\end{tabular}
}

\newpage

\section{Language Modeling Predicates}

\subsection{Language Modeling}

{
\scriptsize \tt  
\begin{tabular}{|ll|} \hline
\multicolumn{2}{|c|}{Preprocessing} \\ \hline
\multicolumn{2}{|l|}{INSERT INTO BASE\_TF(tid, token, tf)} \\
SELECT   & T.tid, T.token, COUNT(*) \\
FROM     & BASE\_TOKENS T \\
GROUP BY & T.tid, T.token\\
& \\
\multicolumn{2}{|l|}{INSERT INTO BASE\_DL(tid, dl)} \\
SELECT   & T.tid, COUNT(*) \\
FROM     & BASE\_TOKENS T \\
GROUP BY & T.TI\\
& \\
\multicolumn{2}{|l|}{INSERT INTO BASE\_PML(tid, token, pml)} \\
SELECT  & T.tid, T.token, T.tf/D.dl \\
FROM    & BASE\_TF T, BASE\_DL D \\
WHERE   & T.tid=D.tid \\
& \\
\multicolumn{2}{|l|}{INSERT INTO BASE\_PAVG(tid, token, pavg)} \\
SELECT  & P.token, AVG(P.pml) \\
FROM    & BASE\_PML P \\
GROUP BY & P.token\\
& \\
\multicolumn{2}{|l|}{INSERT INTO BASE\_FREQ(tid, token, freq)} \\
SELECT  & T.tid, T.token, P.pavg*D.dl \\
FROM    & BASE\_TF T, BASE\_PAVG P, BASE\_DL D \\
WHERE   & T.token = P.token AND T.tid=D.tid \\
& \\
\multicolumn{2}{|l|}{INSERT INTO BASE\_RISK(tid, token, risk)} \\
SELECT  & T.tid, T.token, (1.0/(1.0+Q.freq)) * (POWER(Q.freq/(1.0+Q.freq), T.tf)) \\
FROM    & BASE\_TF T, BASE\_FREQ Q \\
WHERE   & T.tid=Q.tid AND T.token = Q.token\\
& \\
\multicolumn{2}{|l|}{INSERT INTO BASE\_TSIZE(size)} \\
SELECT  & COUNT(*) \\
FROM    & BASE\_TOKENS\\
& \\
\multicolumn{2}{|l|}{INSERT INTO BASE\_CFCS(token, cfcs)} \\
SELECT  & T.token, COUNT(*) / S.size \\
FROM    & BASE\_TOKENS T, BASE\_TSIZE S \\
GROUP BY & T.token\\
& \\
\hline
\end{tabular}
}

{
\scriptsize \tt  
\begin{tabular}{|ll|} \hline
\multicolumn{2}{|l|}{INSERT INTO BASE\_PM(tid, token, pm)} \\
SELECT  & T.tid, T.token, 1E0* POWER(M.pml, 1.0-R.risk) * POWER(A.pavg, R.risk), C.cfcs \\
FROM    & BASE\_TF T, BASE\_RISK R, BASE\_PML M, BASE\_PAVG A, BASE\_CFCS C, BASE\_TSIZE S \\
WHERE   & T.tid=R.tid AND T.token = R.token AND T.tid=M.tid AND\\
        & T.token = M.token AND T.token = A.token AND T.token = C.token\\
& \\
\multicolumn{2}{|l|}{INSERT INTO BASE\_SUMCOMPMBASE(tid, sumcompm)} \\
SELECT  & P.tid, SUM(LOG(1.0-P.pm)) \\
FROM    & BASE\_PM P \\
GROUP BY  & P.tid\\ \hline
\multicolumn{2}{|c|}{Query} \\ \hline
\multicolumn{2}{|l|}{INSERT INTO LM\_RESULTS(tid, score)} \\
SELECT & B1.tid, B1.score + B2.sumcompm \\
FROM   & (SELECT P1.tid, SUM(LOG(P1.pm)) - SUM(LOG(1.0-P1.pm)) - SUM(LOG(P1.cfcs)) AS score \\
       & \hspace{0.03in} FROM BASE\_PM P1, QUERY\_TOKENS T2 \\
       & \hspace{0.03in} WHERE P1.token = T2.token \\
       & \hspace{0.03in} GROUP BY P1.tid) B1, \\
       & BASE\_SUMCOMPMBASE B2 \\
WHERE  & B1.tid=B2.tid \\ \hline
\end{tabular}
}

\newpage

\subsection{Hidden Markov Models}

{
\scriptsize \tt  
\begin{tabular}{|ll|} \hline
\multicolumn{2}{|c|}{Preprocessing} \\ \hline
\multicolumn{2}{|l|}{INSERT INTO BASE\_SIZE(size) } \\
SELECT & COUNT(*) \\
FROM   & BASE\_TABLE\\
& \\
\multicolumn{2}{|l|}{INSERT INTO BASE\_TF(tid, tf)} \\
SELECT   & T.tid, T.token, COUNT(*) \\
FROM     & BASE\_TOKENS T \\
GROUP BY & T.tid, T.token\\
& \\
\multicolumn{2}{|l|}{INSERT INTO BASE\_DL(tid, dl)} \\
SELECT   & T.tid, COUNT(*) \\
FROM     & BASE\_TOKENS T \\
GROUP BY & T.tid\\
& \\
\multicolumn{2}{|l|}{INSERT INTO BASE\_PML(tid, token, pml)} \\
SELECT   & T.tid, T.token, F.tf/D.dl \\
FROM     & BASE\_TOKENS T, BASE\_TF F, BASE\_DL D \\
WHERE    & F.tid=T.tid AND T.token = F.token AND T.tid=D.tid\\
& \\
\multicolumn{2}{|l|}{INSERT INTO BASE\_SUMDL(sdl)} \\
SELECT   & SUM(T.dl) \\
FROM     & BASE\_DL T \\
& \\
\multicolumn{2}{|l|}{INSERT INTO BASE\_PTGE(token, ptge)} \\
SELECT   & T.token, SUM(T.tf)/D.sdl \\
FROM     & BASE\_TF T, BASE\_SUMDL D \\
GROUP BY & T.token\\
& \\
\multicolumn{2}{|l|}{INSERT INTO BASE\_WEIGHTSHMM(tid, token, weight)} \\
SELECT   & M.tid, M.token, LOG( (1 + ($a_1$*M.pml) / ($a_2$*P.ptge)) ) \\
FROM     & BASE\_PTGE P, BASE\_PML M \\
WHERE    & P.token = M.token \\
GROUP BY & tid,token \\
\hline
\multicolumn{2}{|c|}{Query} \\ \hline
\multicolumn{2}{|l|}{INSERT INTO HMM\_SCORES(tid, score)} \\
SELECT   &  W1.tid, EXP(SUM(W1.weight)) \\
FROM     &  BASE\_WEIGHTS W1, QUERY\_TOKENS T2 \\
WHERE    &  W1.token = T2.token \\
\multicolumn{2}{|l|}{GROUP BY  W1.tid} \\ \hline 
\end{tabular}
}

\newpage

\section{Combination Predicates}

\subsection{GES$^{Jaccard}$}

{
\scriptsize \tt  
\begin{tabular}{|ll|} \hline
\multicolumn{2}{|c|}{Preprocessing} \\ \hline
\multicolumn{2}{|l|}{INSERT INTO BASE\_SIZE(size)} \\
SELECT   & COUNT(*) \\
FROM     & BASE\_TABLE\\
& \\
\multicolumn{2}{|l|}{INSERT INTO BASE\_IDF(token,idf)} \\
SELECT   & T.token, LOG(S.size) - LOG(COUNT(DISTINCT T.tid)) \\
FROM     & BASE\_TOKENS T, BASE\_SIZE S \\
GROUP BY & T.token\\
& \\
\multicolumn{2}{|l|}{INSERT INTO BASE\_IDFAVG(idfavg)} \\
SELECT   & AVG(I.idf) \\
FROM     & BASE\_IDF I \\
& \\
\multicolumn{2}{|l|}{INSERT INTO BASE\_TOKENSIZE(tid, token, size)} \\
SELECT   & T.tid, T.token, COUNT(*)\\
FROM     & BASE\_QGRAMS T \\
GROUP BY & T.tid, T.TOKE\\
& \\
\multicolumn{2}{|l|}{INSERT INTO BASE\_QGRAMSTOKENSIZE(tid, token, qgram, size)}\\
SELECT   & T.tid, T.token, T.qgram, S.size \\
FROM     & BASE\_QGRAMS T, BASE\_TOKENSIZE S \\
WHERE    & T.tid = S.tid AND T.token = S.token\\
\hline
\end{tabular}
}

{
\scriptsize \tt  
\begin{tabular}{|ll|} \hline
\multicolumn{2}{|c|}{Query (Filtering Step)} \\ \hline
SELECT 	 & MAXSIM.tid, R.string \\
       	 & (1.0 - 1.0/$q$) + (1/SUM(QIDF.idf)) * SUM(QIDF.idf * (((2.0/$q$) * MAXSIM.maxsim) )) AS score \\
FROM 	 & ( SELECT JAC\_SIM.tid, JAC\_SIM.token2, MAX(sim) AS maxsim \\
     	 &   FROM (SELECT BSIZE.tid AS tid, BSIZE.token AS token1, Q.token AS token2, \\
     	 &   \hspace{0.68in}     COUNT(*)/(BSIZE.size + QSIZE.size - COUNT(*)) AS sim \\
     	 &   \hspace{0.28in}      FROM BASE\_QGRAMSTOKENSIZE BSIZE, QUERY\_QGRAMS Q, \\
     	 &   \hspace{0.68in}           (SELECT T.token, COUNT(*) AS size \\
     	 &   \hspace{0.68in}            FROM QUERY\_QGRAMS T \\
     	 &   \hspace{0.68in}            GROUP BY T.token) QSIZE \\
     	 &   \hspace{0.28in}      WHERE BSIZE.qgram = Q.qgram AND Q.token = QSIZE.token \\
     	 &   \hspace{0.28in}      GROUP BY BSIZE.tid, BSIZE.token, Q.token) JAC\_SIM \\
     	 &   GROUP BY JAC\_SIM.tid, JAC\_SIM.token2 ) MAXSIM, \\
     	 &  (SELECT R.token, R.idf \\
		 &   FROM QUERY\_TOKENS S, BASE\_IDF R\\
		 &   WHERE S.token = R.token \\
		 &   GROUP BY S.token \\
		 &   UNION \\
		 &   SELECT S.token, A.IDFAVG AS idf \\
		 &   FROM QUERY\_TOKENS S, BASE\_IDFAVG A \\
		 &   WHERE S.token NOT IN (SELECT I.token FROM BASE\_IDF I) \\
		 &   GROUP BY S.token) QIDF, \\
         &  BASE\_TABLE R \\
WHERE    &  TM.token2 = I.token AND R.tid = MAXSIM.tid \\
GROUP BY &  TM.tid \\
HAVING   &  score >= $\theta$ \\ \hline
\end{tabular}
}

\newpage

\subsection{GES$^{apx}$}

{
\scriptsize \tt  
\begin{tabular}{|ll|} \hline
\multicolumn{2}{|c|}{Preprocessing} \\ \hline
\multicolumn{2}{|l|}{INSERT INTO BASE\_SIZE(size)} \\
SELECT   & COUNT(*) \\
FROM     & BASE\_TABLE\\
& \\
\multicolumn{2}{|l|}{INSERT INTO BASE\_IDF(token, size)} \\
SELECT   & T.token, LOG(S.size) - LOG(COUNT(DISTINCT T.tid)) \\
FROM     & BASE\_TOKENS T, BASE\_SIZE S \\
GROUP BY & T.token\\
& \\
\multicolumn{2}{|l|}{INSERT INTO BASE\_IDFAVG(idfavg)} \\
SELECT   & AVG(I.idf) \\
FROM     & BASE\_IDF I \\
& \\
\multicolumn{2}{|l|}{INSERT INTO BASE\_HASHFUNC(fid, func)} \\
SELECT   & N.i-1, round(rand()*MAXINT) \\
FROM     & INTEGERS N \\
LIMIT    & HASH\_SIZE \\
& \\
\multicolumn{2}{|l|}{INSERT INTO BASE\_HASHVALUE(fid, qgram, value)}  \\
SELECT   & F.FID, Q.QGRAM, MOD(CONV(HEX( Q.qgram), 16, 10) * MAXINT, F.func)  \\
FROM     & BASE\_HASHFUNC F, (SELECT DISTINCT QGRAM FROM BASE\_QGRAMS) Q \\
& \\
\multicolumn{2}{|l|}{INSERT INTO BASE\_MINHASHSIGNATURE(tid, token, fid, score)} \\
SELECT   & Q.tid, Q.token, H.fid, MIN(H.value) \\
FROM     & BASE\_QGRAMS Q, BASE\_HASHVALUE H \\
WHERE    & Q.QGRAM = H.QGRAM \\
GROUP BY & Q.tid, Q.token, H.fid\\
\hline
\end{tabular}
}

{
\scriptsize \tt  
\begin{tabular}{|ll|} \hline
\multicolumn{2}{|c|}{Query (Filtering Step)} \\ \hline
SELECT   & MAXSIM.tid, R.string, \\
     	 & (1.0 - 1.0/$q$) + (1/SUM(I.idf)) * SUM(I.idf * (((2.0/$q$) * MAXSIM.maxsim) )) AS score \\
FROM 	 & (SELECT MH\_SIM.tid, MH\_SIM.token2, MAX(sim) AS maxsim \\
      	 & FROM (SELECT BMHSIG.tid AS tid, BMHSIG.token AS token1, \\
         & \hspace{0.70in}  QMHSIG.token AS token2, COUNT(*)/$H$ AS sim \\
      	 & \hspace{0.28in}      FROM BASE\_MINHASHSIGNATURE BMHSIG, \\
      	 & \hspace{0.68in}           (SELECT Q.token, H.fid, MIN(H.value) AS value \\
      	 & \hspace{0.68in}            FROM QUERY\_QGRAMS Q, BASE\_HASHVALUE H \\
      	 & \hspace{0.68in}            WHERE Q.qgram = H.qgram \\
      	 & \hspace{0.68in}            GROUP BY Q.token, H.fid) QMHSIG \\
      	 & \hspace{0.28in}      WHERE BMHSIG.fid = QMHSIG.fid AND BMHSIG.value = QMHSIG.value \\
      	 & \hspace{0.28in}      GROUP BY BMHSIG.tid, BMHSIG.token, QMHSIG.token) MH\_SIM \\
      	 & GROUP BY MH\_SIM.tid, MH\_SIM.token2) MAXSIM, \\
     	 &  (SELECT R.token, R.idf \\
		 &   FROM QUERY\_TOKENS Q, BASE\_IDF R\\
		 &   WHERE Q.token = R.token \\
		 &   GROUP BY Q.token \\
		 &   UNION \\
		 &   SELECT Q.token, A.IDFAVG AS idf \\
		 &   FROM QUERY\_TOKENS Q, BASE\_IDFAVG A \\
		 &   WHERE Q.token NOT IN (SELECT I.token FROM BASE\_IDF I) \\
		 &   GROUP BY Q.token) QIDF, \\
         & BASE\_TABLE R \\      	 
WHERE    & MAXSIM.token2 = I.token AND R.tid = MAXSIM.tid \\
GROUP BY & MAXSIM.tid \\
HAVING   & score >= $\theta$ \\
\hline
\end{tabular}
}

\newpage

\subsection{SoftTFIDF}

{
\scriptsize \tt  
\begin{tabular}{|ll|} \hline
\multicolumn{2}{|c|}{Preprocessing} \\ \hline
\multicolumn{2}{|l|}{INSERT INTO BASE\_SIZE(size)} \\
SELECT   & COUNT(*) \\
FROM     & BASE\_TABLE\\
& \\
\multicolumn{2}{|l|}{INSERT INTO BASE\_IDF(token, idf)} \\
SELECT   & T.token, LOG(S.size) - LOG(COUNT(DISTINCT T.tid)) \\
FROM     & BASE\_TOKENS T, BASE\_SIZE S \\
GROUP BY & T.token\\
& \\
\multicolumn{2}{|l|}{INSERT INTO BASE\_TF(tid, token, tf)} \\
SELECT   & T.tid, T.token, COUNT(*) \\
FROM     & BASE\_TOKENS T \\
GROUP BY & T.tid, T.token\\
& \\
\multicolumn{2}{|l|}{INSERT INTO BASE\_LENGTH(tid, len)} \\
SELECT   & T.tid, SQRT(SUM(I.idf*I.idf*T.tf*T.tf)) \\
FROM     & BASE\_IDF I, BASE\_TF T \\
WHERE    & I.token = T.token \\
GROUP BY & T.tid\\
& \\
\multicolumn{2}{|l|}{INSERT INTO BASE\_WEIGHTS(tid, token, weight))} \\
SELECT   & T.tid, T.token, I.idf*T.tf/L.len \\
FROM     & BASE\_IDF I, BASE\_TF T, BASE\_LENGTH L \\
WHERE    & I.token = T.token AND T.tid = L.tid \\
\hline
\end{tabular}
}

{
\scriptsize \tt  
\begin{tabular}{|ll|} \hline

\multicolumn{2}{|c|}{Query} \\ \hline
\multicolumn{2}{|l|}{INSERT INTO SOFTTFIDF\_RESULTS(tid, score)} \\
SELECT   &  MAXTOKEN.tid, SUM(WB.weight * WQ.weight * MAXTOKEN.maxsim) \\
FROM 	 &  BASE\_WEIGHTS WB,\\
     	 &  (SELECT JARO\_SIM.tid, JARO\_SIM.token1, JARO\_SIM.token2, MAXSIM.maxsim \\
     	 &  FROM           (SELECT JARO\_SIM.tid, JARO\_SIM.token2, MAX(sim) AS maxsim \\
     	 &  \hspace{0.30in} FROM \hspace{0.07in}(SELECT R1.tid AS tid, R1.token AS token1, \\
         &  \hspace{1.14in}                             R2.token AS token2, JaroWinkler(R1.token,R2.token) AS sim \\
     	 &  \hspace{0.68in}                      FROM BASE\_TOKENS R1, QUERY\_TOKENS R2 \\
     	 &  \hspace{0.68in}                      WHERE JaroWinkler(R1.token,R2.token) >= $\theta$) JARO\_SIM \\
     	 &  \hspace{0.30in} GROUP BY JARO\_SIM.tid, JARO\_SIM.token2) MAXSIM, \\
     	 &  \hspace{0.30in}(SELECT R1.tid AS tid, R1.token AS token1, \\
         &  \hspace{0.68in}        R2.token AS token2, JaroWinkler(R1.token,R2.token) AS sim \\
     	 &  \hspace{0.30in} FROM BASE\_TOKENS R1, QUERY\_TOKENS R2 \\
     	 &  \hspace{0.30in} WHERE JaroWinkler(R1.token,R2.token) >= $\theta$) JARO\_SIM \\
     	 &  WHERE  JARO\_SIM.tid = MAXSIM.tid AND JARO\_SIM.token2 = MAXSIM.token2 \\
         &  \hspace{0.30in} AND MAXSIM.maxsim = JARO\_SIM.sim) MAXTOKEN, \\
     	 & (SELECT QTF.token, QIDF.idf*QTF.tf/QLEN.length AS weight \\
     	 &  FROM (SELECT R.token, R.idf \\
     	 &  \hspace{0.30in}      FROM QUERY\_TOKENS S, BASE\_IDF R \\
     	 &  \hspace{0.28in}      WHERE S.token = R.token \\
     	 &  \hspace{0.28in}      GROUP BY S.token) QIDF, \\
     	 &  \hspace{0.28in}     (SELECT T.token, COUNT(*) AS tf \\
     	 &  \hspace{0.28in}      FROM QUERY\_TOKENS T \\
     	 &  \hspace{0.28in}      GROUP BY T.token) QTF, \\
     	 &  \hspace{0.28in}     (SELECT QTF.tid, SQRT(SUM(QIDF.idf*QIDF.idf*QTF.tf*QTF.tf)) AS length \\
     	 &  \hspace{0.28in}      FROM (SELECT R.token, R.idf \\
     	 &  \hspace{0.68in}            FROM QUERY\_TOKENS S, BASE\_IDF R \\
     	 &  \hspace{0.68in}            WHERE S.token = R.token \\
     	 &  \hspace{0.68in}            GROUP BY S.token) QIDF, \\
     	 &  \hspace{0.68in}           (SELECT T.token, COUNT(*) AS tf \\
     	 &  \hspace{0.68in}            FROM QUERY\_TOKENS T \\
     	 &  \hspace{0.68in}            GROUP BY T.token) QTF \\
     	 &  \hspace{0.28in}      WHERE QIDF.token = QTF.token) QLEN \\
     	 &  WHERE QIDF.token = T.token) WQ, \\
WHERE    & MAXTOKEN.token2 = WQ.token AND MAXTOKEN.tid = WB.tid AND MAXTOKEN.token1 = WB.token \\
GROUP BY & MAXTOKEN.tid\\ 
\hline
\end{tabular}
}

\end{document}